\documentclass[12pt,preprint]{aastex}

\newcommand\kms{$\rm km\,s^{-1}$}
\newcommand\asec{^{\prime\prime}}
\newcommand\Msun{{\rm M}_{\odot}}
\newcommand\ea{et al.~}

\newcommand\Tlight{$\overline{T}_{light}$}
\newcommand\Tmass{$\overline{T}_{mass}$}

\shorttitle{Bulges Central Stellar Populations}
\shortauthors{Sarzi et al.}

\begin{document}

%--- TITLE
\title{The Stellar Populations in the Central Parsecs of Galactic Bulges}

\footnotetext[1]{Based on observations obtained with the {\it Hubble Space
Telescope}, which is operated by AURA, Inc., under NASA contract NAS5-26555.}

\author{
Marc~Sarzi\altaffilmark{2}, Hans-Walter~Rix\altaffilmark{3},
Joseph~C.~Shields\altaffilmark{4}, Luis~C.~Ho\altaffilmark{5},
Aaron~J.~Barth\altaffilmark{6}, Gregory~Rudnick\altaffilmark{7},
Alexei~V.~Filippenko\altaffilmark{8}, and
Wallace~L.~W.~Sargent\altaffilmark{9} }

\altaffiltext{2}{University of Oxford, Astrophysics, Keble Road, 
  OX13RH Oxford, United Kingdom; sarzi@astro.ox.ac.uk}
\altaffiltext{3}{Max-Planck-Institut f\"ur Astronomie, K\"onigstuhl 17, 
  D-69117 Heidelberg, Germany; rix@mpia-hd.mpg.de}
\altaffiltext{4}{Physics \& Astronomy Department, Ohio University,
  Athens, OH 45701; shields@phy.ohiou.edu}
\altaffiltext{5}{The Observatories of the Carnegie Institution of
  Washington, 813 Santa Barbara St., Pasadena, CA 91101-1292;
  lho@ociw.edu}
\altaffiltext{6}{Department of Physics \& Astronomy, 4129 Frederick 
 Reines Hall, University of California, Irvine, CA 92697-4575;
 barth@uci.edu}
\altaffiltext{7}{Max-Planck-Institut f\"ur Astrophysik, 
  Karl-Schwarzschild-Strasse 1, D-85741 Garching, Germany; 
  grudnick@mpa-garching.mpg.de}
\altaffiltext{8}{Department of Astronomy, University of California,
  Berkeley, CA 94720-3411; alex@astro.berkeley.edu}
\altaffiltext{9}{Astronomy Department, California Institute of Technology, 
  MS 105-24, Pasadena, CA 91125; wws@astro.caltech.edu}

\begin{abstract}
We present {\it Hubble Space Telescope\/} blue spectra at intermediate
spectral resolution for the nuclei of 23 nearby disk galaxies.  These
objects were selected to have nebular emission in their nuclei, and
span a range of emission-line classifications as well as Hubble types.
In this paper we focus on the stellar population as revealed by the
continuum spectral energy distribution measured within the central
$0\farcs13$ ($\sim 8$ pc) of these galaxies.
The data were modeled with linear combinations of single-age stellar
population synthesis models.
The large majority ($\sim 80$\%) of the surveyed nuclei have spectra
whose features are consistent with a predominantly old ($\gtrsim 5
\times 10^9$ yr) stellar population. Approximately 25\% of these
nuclei show evidence of a component with age younger than 1 Gyr, with
the incidence of these stars related to the nebular classification.
Successful model fits imply an average reddening corresponding to $A_V
\approx 0.4$ mag and a stellar metallicity of (1--2.5)$Z_\odot$.
We discuss the implications of these results for the understanding of
the star formation history in the environment of quiescent and active
supermassive black holes.
Our findings reinforce the picture wherein Seyfert nuclei and the
majority of low-ionization nuclear emission-line regions (LINERs) are
predominantly accretion-powered, and suggest that much of the central
star formation in \ion{H}{2} nuclei is actually circumnuclear.

\end{abstract}


%--- INTRODUCTION
\section{Introduction}
\label{sec:Pops_intro}

The centers of galaxies hold special interest as the locations of
unusual energetic phenomena.
It is now commonly accepted that most galaxies harbor central
supermassive black holes (SMBHs) that in the past may have shone as
powerful active galactic nuclei (AGNs). The fueling of luminous AGNs
requires substantial gas concentrations that might also be favorable
sites for star formation, and the formation of a dense star cluster is
in fact one means of spawning a SMBH \citep[e.g.,][]{Ree84}.
In the present Universe, many galactic nuclei exhibit emission-line
activity at low levels that apparently derives from a variable mix of
accretion power and star formation \citep[e.g.,][]{Ho97}.
Emission-line ratios provide a basis for classifying objects into
standard categories that, to varying degrees of confidence, are
identified with specific power sources. While classification as a
Seyfert or \ion{H}{2} nucleus is routinely associated with nonstellar
accretion and O stars as dominant energy sources, respectively, many
galaxies are classified as low-ionization nuclear emission-line
regions \citep[LINERs]{Heck80} or LINER/\ion{H}{2} ``transition
nuclei'' \citep[e.g.,][]{Fil92,Ho93} which are more ambiguous in their
interpretation.  Evidence increasingly suggests that LINERs are
primarily accretion-powered \citep[e.g.,][]{Ho03,Fil03}, but the situation
is less clear for the transition sources.

The direct detection of young stellar populations in galactic nuclei
is an important means of quantifying star formation in these
environments and possible connections to emission-line activity.
Nearby examples suggest that recent star formation is not unusual.
The Milky Way, IC~342, and NGC~4449 all have central young stellar
clusters \citep{Kra95,Boe01,Boe99}, and our neighbors M33 and M31 also
have blue nuclei that are most likely consistent with the presence of
young stars \citep{Lau98}.
Moreover, imaging surveys with the {\it Hubble Space Telescope\/} ({\it 
HST\/}) have shown that a compact star cluster is often present in the
center of spiral galaxies \citep{Phi96,Caro98,Mat99}, in particular in
the latest Hubble types \citep{Boe02}.
If proved young, such clusters may help account for the high frequency
(41\%) of \ion{H}{2} nuclei in nearby galaxies \citep{Ho97}, and may
also have bearing on transition nuclei.

Most existing studies of stellar populations and nebular emission in
galaxy centers have been carried out from the ground, and in probing
``nuclei'' are thus limited to studying regions that may actually be
several hundreds of parsecs in scale. Considerable structure may exist
on smaller scales, and the spatial distribution of star formation
within such a region is of particular interest.
A number of ultraviolet (UV) {\it HST\/} surveys have shown that star
formation in galactic nuclei often takes place in circumnuclear rings
\citep[e.g.,][]{Mao96}.
If all bulges harbor a SMBH, there are reasons to believe that it may
be difficult to form stars in their very central regions. In
particular, the tidal field induced by a SMBH may suffice to disrupt
molecular clouds before they can gravitationally collapse.
Observationally, however, a number of studies have shown that SMBHs
and young stellar clusters can coexist on very small scales, at least
in some Seyfert~2 and LINER nuclei \citep{Heck97,Gonz98,Col02} and in
the Milky Way \citep[e.g.,][]{Gen03}.

Probing galaxy centers with the highest possible spatial resolution is
thus of great value for understanding what powers ``nuclear''
activity, and how star formation proceeds in these zones.
The recent {\it HST\/} surveys of \citet{Pel99} and
\citet{Caro01,Caro02} represent excellent examples of how
high-resolution images have already helped in making steps in this
direction.
The observed visual to near-infrared colors indicate that the vast
majority of disk galaxy centers are very red, consistent with
intermediate to very old stellar populations that are substantially
obscured by interstellar material.
Yet, with only broad-band colors it is difficult to disentangle the
effects of an old stellar population from those of reddening by dust,
super-solar metallicity, or an active nucleus.

For this reason we undertook a survey of nearby nuclei with the Space
Telescope Imaging Spectrograph (STIS) on board {\it HST\/} (the SUNNS
survey), acquiring moderate-resolution blue and red spectra for 23
galaxies to study both their nuclear stellar populations and
emission-line kinematics, and to detect in particular the presence of
very young stars or SMBHs.
In this paper we present a study of the nuclear stellar populations of
the SUNNS sample galaxies, based on the fitting of synthetic stellar
population models to the observed blue spectra.  After we submitted
this paper for publication, we learned of related work by
\citet{Gonz04} which reaches broadly similar conclusions.

The paper is organized as follows. In \S 2 we show the nuclear spectra
for our sample objects, in \S 3 we describe our analysis method, and
in \S 4 we present our results. Our analysis includes models based on
single-age stellar populations (\S 4.1) or with extended star
formation histories (\S 4.2) and addresses the effect of super-solar
metallicities (\S 4.3). From the model fitting we derive conservative
upper limits on the presence of very young stars and constraints on
the need for absorption-line dilution by a featureless continuum (\S
4.4).  We discuss these results in \S 5 and draw our
conclusions in \S 6.

%--- OBSERVATIONS AND DATA REDUCTION
\section{Observations and Data Reduction}
\label{sec:Pops_obs}

The nuclear spectral energy distributions (SEDs) modelled in this
paper have been extracted from the two-dimensional (2-D) {\it
HST\/}-STIS spectra acquired in the course of the SUNNS spectroscopic
survey of 23 galactic nuclei \citep{Shi04b}. The
SUNNS targets were drawn from the Palomar spectroscopic survey of
nearby galaxies \citep{Fil85,Ho97} and include all galaxies within
17~Mpc known to have optical H$\alpha$ or [\ion{N}{2}] $\lambda$6583
line emission ($\gtrsim 10^{15}$ ergs s$^{-1}$ cm$^{-2}$) within a
$2\asec \times 4\asec$ aperture. The sample was also chosen to
investigate the nature of nuclear activity and includes Seyfert
galaxies, LINERs, LINER/\ion{H}{2} transition objects, and \ion{H}{2}
nuclei.
Basic parameters of the sample galaxies are given in Table
\ref{tab:Pops_GalSample}.

Spectra were acquired with the STIS $0\farcs2 \times 50\asec$ slit,
and 5-pixel-wide ($\sim 0\farcs 25$) extractions were used to generate
one-dimensional (1-D) spectra of the nuclei.  The resulting aperture
of $0\farcs2 \times 0\farcs25$ is equivalent in area to a circular
aperture with radius $r=0\farcs13$, corresponding to 8.2 pc for the
mean sample distance of $\sim 13$ Mpc. The G430L and G750M gratings
were employed to obtain spectral coverage of 3000--5700~\AA\ and
6300--6850~\AA, respectively. In the present analysis we utilize only
the G430L spectra, since these data cover spectral features providing
the strongest constraints on the stellar population.
As extinction by interstellar dust will be an important factor in our
modelling, we have corrected the observed fluxes for Galactic
foreground reddening, using $A_V$ values from \citet{Schl98} and the
\citet{Card89} extinction law with $R_{\rm V} = 3.1$.
The full-width at half maximum (FWHM) resolution for the G430L spectra
is 7.8~\AA\ for extended sources. Full details concerning the
observations are presented by \citet{Shi04b}.

\begin{figure}\plotone{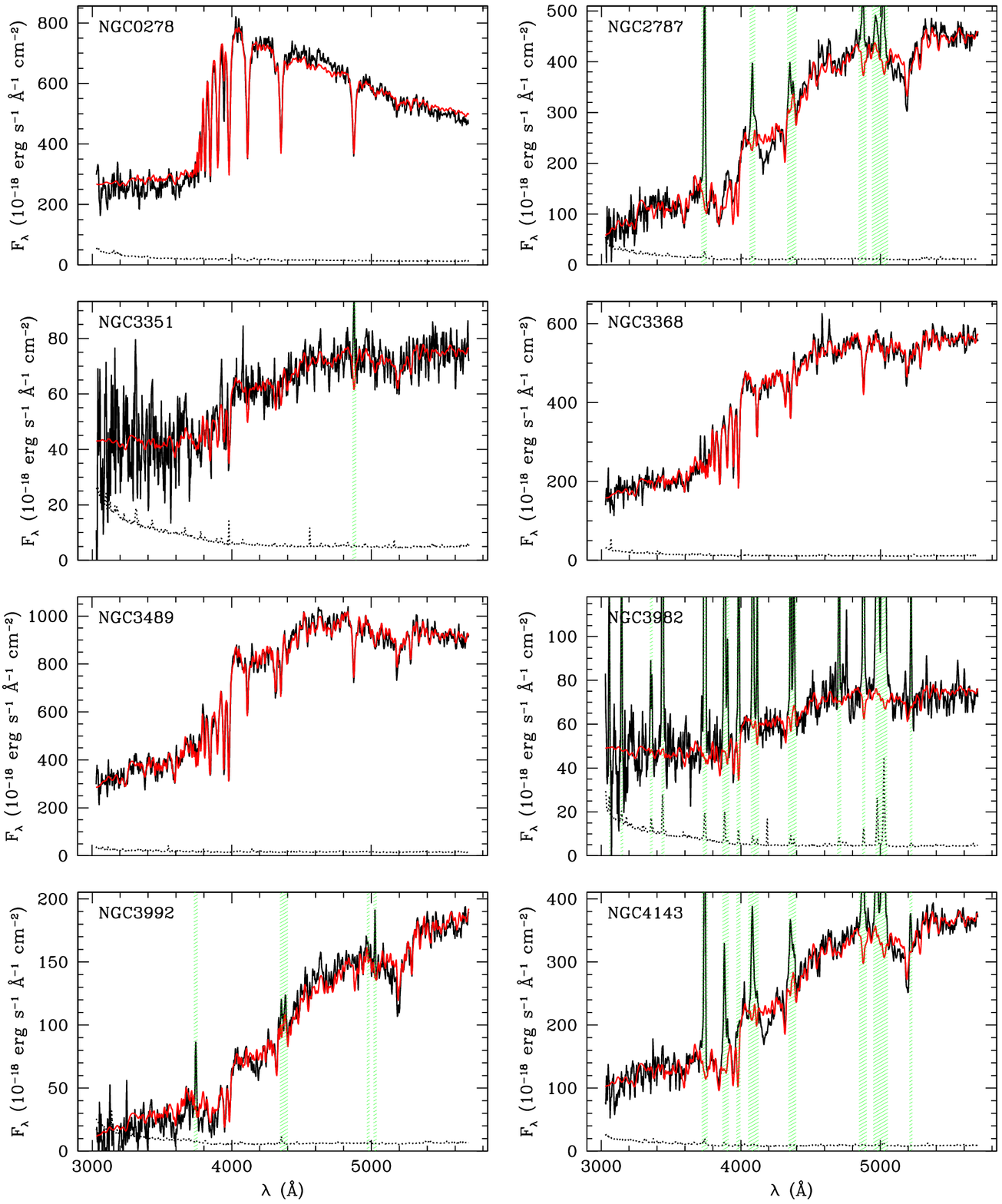}\caption{
 Nuclear G430L spectra extracted from the central $0\farcs25 \times
 0\farcs2$ of our sample galaxies. The {\it dotted lines} at the
 bottom of each panel show the error array corresponding to the
 observed fluxes, while the {\it dashed green columns} indicate the
 spectral regions excluded during our analysis in order to avoid
 emission lines. The {\it red lines} show our best-fitting
 multiple-starburst population models. The observed fluxes have been
 corrected for Galactic foreground extinction, while the models
 account for dust extinction intrinsic to the
 source.\label{fig:Pops_data_and_magefit}
 }\end{figure}

The final extracted and dereddened nuclear G430L spectra, along with
their associated errors, are shown for all our sample galaxies in
Figure \ref{fig:Pops_data_and_magefit}. Their heterogeneity is clear,
in terms of both their quality and the kind of spectral energy
distribution they are sampling. The signal-to-noise ratio
(SNR) per pixel is $\sim$2--15 and $\sim$15--50 at the blue and red ends,
respectively.
Most nuclear spectra have a strong 4000~\AA\ break, as well as Mg and
Fe absorption features (around 5175~\AA\ and 5270~\AA, respectively),
indicative of old stellar populations (e.g., NGC~3992, NGC~4459, or
NGC~4596), but some show strong Balmer absorption lines underscoring
the presence of younger components (e.g., NGC~3368 and NGC~3489).
Many nuclei display obvious nebular emission, which must be excluded
when analyzing the underlying stellar spectrum.
Contamination by line emission is particularly strong for the Seyfert
galaxy NGC~3982 and the two LINERs NGC~4203 and NGC~4450, where a very
broad pedestal of Balmer emission has already been reported by
\citet{Shi00} and \citet{Ho00}, respectively.
Figure \ref{fig:Pops_data_and_magefit} shows for each galaxy the
spectral regions omitted for this reason in our analysis.

\subsection{Bulge-Light Contamination}
\label{subsec:Pops_obs_bulgecontamin}

An important issue is the extent to which our projected nuclear
spectra sample the actual nuclear regions of our galaxies.
To address this matter, we derived for each object an estimate for the
intrinsic luminosity density profile, which can be used to determine
the ratio between the light emitted within the central sphere of
radius equal to our central aperture, $r\approx 0\farcs13$, and the
light collected along the line of sight in a cylindrical beam with the
same radius.
The luminosity density profile was estimated by assuming spherical
symmetry and deprojecting the stellar surface brightness distribution
in the STIS acquisition image.  The multi-Gaussian algorithm adopted
for this purpose by \citet{Sar01} was applied to circularly averaged
surface brightness profiles extracted using the {\scriptsize
IRAF}\footnote{IRAF is distributed by the National Optical Astronomy
Observatories, which are operated by the Association of Universities
for Research in Astronomy, Inc., under cooperative agreement with the
National Science Foundation.} task {\scriptsize ELLIPSE}.
Dust absorption causes problems for this approach in some cases, but
the analysis overall indicates that typically $\sim$60\% of the flux
collected within our central aperture should come from regions within
$\sim 8$ pc of the center.

As our sample consists of relatively early-type disk galaxies (S0 to
Sb) with intermediate inclinations (average inclination angle
44$^{\circ}$, Table \ref{tab:Pops_GalSample}), contamination of our
spectra by extranuclear stars seen in projection is presumably due
mostly to bulge stars, rather than stars in the galactic disk.

\placefigure{fig:Pops_data_and_magefit} 
\notetoeditor{Figure 1 should occupy three pages when printed, each
part being on top of the page and being as wide as it}

%--- THE RIX \& WHITE METHOD
\section{Analysis Method}
\label{sec:Pops_method}

To interpret our nuclear spectra, we fit them with synthetic
population models that are based on a variety of star formation
histories, metallicities, and reddenings.  To simulate extended star
formation histories, we used linear combinations of single-age stellar
population synthesis templates to fit the nuclear spectra directly in
pixel space, following \citet{Rix95}.
This algorithm derives simultaneously the optimal combination of
template spectra and the kinematical broadening required to match the
data.
To establish the relative weight of different model SEDs in explaining
each galaxy nuclear spectrum, we proceeded through the following
steps.

\placefigure{fig:Pops_BC03}
\notetoeditor{Figure 2 should be on top of the page and occupy all its width}

\begin{enumerate}

\item Select one template from the set of model SEDs and use it to obtain 
      an initial estimate for the line-of-sight velocity distribution
      (LOSVD), which here in practice consists of a simple Gaussian.

\item Use the LOSVD obtained in step (1) to shift and broaden 
      {\em all} of the model SEDs, and find the linear combination
      that best matches the observed spectrum, using a non-negative
      least-squares algorithm.  This prescription assumes that
      different stellar populations in a given nucleus share the same
      kinematics.

\item Use the linear combination of weights obtained in step (2) and the 
      original model SEDs to construct an unbroadened optimal
      template, which is appropriate for matching the strength of
      spectral features in the observed spectrum.

\item Repeat steps (1), (2), and (3) using this first optimal template in order
      to derive an improved LOSVD, model SED weights, and optimal
      template.

\end{enumerate}

By working in pixel space rather than Fourier space, it is
straightforward in steps (2) and (3) to exclude wavelength intervals that
are contaminated by emission lines. Inspection of the fit residuals
enable a better location of such features.

Our library of template spectra is based on the Bruzual \& Charlot
(2003, hereafter B\&C) model SEDs for single-age stellar populations
with a \citet{Sal55} initial mass function, with 11 different ages
$T_i$ spanning 1 Myr to 10 Gyr.  The specific ages employed are
$\log{T_i}= 6.0, 6.5, 7.0, 7.5, 8.0, 8.5, 9.0, 9.25, 9.50, 9.75, {\rm
and}\, 10.0$, where $T_i$ is expressed in years.
In our analysis we use exclusively B\&C model SEDs with solar and 2.5
times solar metallicity.  These models have spectral resolution (FWHM
= 3~\AA) higher than that of our observations, and thus are well
suited for obtaining detailed information from spectral features in
the data.  Figure \ref{fig:Pops_BC03} shows the single-starburst
population (SSP) models adopted for our analysis.

In our fit analysis, we included the possibility of interstellar
reddening intrinsic to the source, which we modeled using sets of B\&C
templates modified by different $A_V$ values and the \citet{Card89}
extinction law.
We considered fractions of absorbed flux at 5512~\AA\ amounting to 9,
17, 24, 30, 37, 42, 48, 52, 56, and 60\%, corresponding to $A_V$ of
0.1, 0.2, 0.3, 0.4, 0.5, 0.6, 0.7, 0.8, 0.9, and 1.0 mag, respectively.

Since the B\&C model SEDs correspond to ``clusters'' of 1 M$_\odot$,
both in stars and the gaseous material ejected by them during their
evolution, the linear fit coefficients are total mass weights
$m_i$. The corresponding stellar mass weights are $m^{\star}_i =
f^{\star}_i m_i$, where $f^{\star}_i$ is the age-dependent
fraction of the total mass which is in stars. The stellar
mass-weighted mean age is then given by $\overline{T}_{mass} = (\sum_i
T_i m^{\star}_i)/ \sum_i m^{\star}_i$ for the best multi-age
models.
Alternatively, if template SEDs are scaled to the same luminosity, the
relative weights $l_i$ reflect the luminosity-weighted mean age
$\overline{T}_{light} = (\sum_i T_is l_i)/ \sum_i l_i$.
Scaling all the 1-M$_\odot$ model SEDs to the same luminosity is
equivalent to dividing them by their total mass-to-light ratio
$\Upsilon$ times a constant, so that the light weights are just $l_i
\propto m_i/\Upsilon_i = m^{\star}_i/\Upsilon^{\star}_i$, where
$\Upsilon^{\star} = \Upsilon f^{\star}_i$ is the stellar
mass-to-light ratio.  It follows that \Tlight\ is always smaller than
\Tmass\ because older SSP models always have larger mass-to-light
ratios than younger ones.
Unless the observed SED is clearly dominated by old stars, the
luminosity-weighted mean age estimate \Tlight\ is more robust than
\Tmass, because any small error in matching the
observed fluxes becomes more relevant for older stars than for younger
stars when it comes to addressing mass fractions.
On the other hand, in the limit of a small component of young stars
intermingled with an old population, a varying contribution of young
stars will shift \Tlight\ much more than \Tmass.
In this situation we are more likely to obtain an upper limit on the
contribution of young stars, and the derived luminosity-weighted mean
age \Tlight\ will still represent a safe lower limit for the age of
the nuclear population.

In the following analysis we therefore focus on the \Tlight\ values,
as robust estimates or lower limits on the stellar population
age. Conversely, the mass fractions of young stars ($\lesssim$ 1 Gyr)
should be regarded as upper limits. We emphasize that these statements
apply to the stellar populations to the extent that they are traced by
optical light.  Our analysis is not sensitive to young populations
subject to large extinction, which in principle could be present but
elude detection in our data.

Alternatively, we could have described the stellar populations of our
sample nuclei through standard Lick/IDS indices, to be used in
diagnostic diagrams \citep[e.g,][]{Wor94,Vaz99}.
Given the considerable amount of nebular emission often present in our
nuclear spectra, the best choice for an age indicator would have been
the H$\gamma_A$ index introduced by \citet{Wor97}; this feature is
less contaminated by gas emission than is H$\beta$, since the nebular
flux decreases rapidly with increasing order for the Balmer
lines. Furthermore, since many of our nuclear spectra are fairly
noisy, as a metallicity indicator we could have considered the Fe3
index of \citet{Kun00}, which combines three strong iron lines and is
also somewhat less sensitive to non-solar metal abundance ratios.
However, the limited spectral resolution of our G430L spectra does not
allow accurate measurement of stellar velocity dispersions from these
data, which thus precludes a corresponding correction of the derived
indices. In particular, the kinematical broadening would reduce the
line strength of the metallicity indicator \citep{Kun00}, which would
bias us to infer older ages.
As the effect is increasingly important for larger velocity
dispersions, this problem would have been particularly serious in the
case of our nuclear spectra, because stellar velocity dispersions are
expected to reach large values in proximity of the central SMBH.
The use of line indices also does not readily allow for the influence
of an AGN continuum, which should be considered given indications that
many of our sample members have an accretion-powered source.
A template superposition approach consequently is the preferred
methodology for interpreting the current data.

%--- RESULTS
\section{Results}
\label{sec:Pops_results}

\subsection{Single-Starburst Population Models}
\label{subsec:Pops_results_sage}

For an initial description of the nuclear stellar population in our
sample galaxies, we determined which of the single-age
solar-metallicity model SEDs considered alone best matched the
observed spectrum for each source.
In our analysis we left the intrinsic $A_V$ as a parameter free to
assume any of the values specified in \S\ref{sec:Pops_method}.
Table \ref{tab:Pops_sage} lists the ages and $A_V$ values obtained
from this fitting exercise, along with the corresponding $\chi^2$
values per degree of freedom ($\chi^2_\nu$) as an indication of the
quality of each model.
Clearly, some spectra are poorly described only by reddened single-age
populations.
Nevertheless, the most immediate and significant finding is that the
populations tend to be old --- the single-age values for most reliable
models exceed $10^{9.5}$ yr $\approx 3$ Gyr. Noticeable exceptions to
this trend are NGC~278, NGC~3351, and NGC~3368, for which a relatively
young population is able to explain most observed features.
On average, the best-fit estimate of internal extinction is modest
($\overline{A}_V = 0.3$ mag), but many of the reddening values are not
reliable in light of problems with the quality of the model fit.

\subsection{Multiple-Starburst Population Models}
\label{subsec:Pops_results_mage}

\subsubsection{Multi-Component Fits}

Since it is very probable that the galaxy nuclei contain composite
stellar populations, we repeated our analysis with linear combinations
of template SEDs drawn from the full set of solar-metallicity SSP
models.  The intrinsic $A_V$ was again left as a free parameter, with
a common reddening assumed for all templates in modeling a given
source.
This approach generates estimates of the luminosity-weighted mean age
and the mass fraction of young stars ($\lesssim$1 Gyr) as described in
\S\ref{sec:Pops_method}, and provides an indication of whether it is 
necessary to invoke multiple or extended star-formation episodes, or
very young stars (few Myr), to explain the observed SEDs.
If two or more templates representing adjacent steps in our age grid
are assigned comparable weights, we typically cannot conclude with
confidence that multiple star-formation events took place, because of
the discrete and rather coarse age sampling of the model templates.
The case for multiple episodes is thus strong only if model components
of very different age are required to match the data.
It may also be difficult to distinguish between contributions from a
few-Myr-old stellar population and from an AGN power law, particularly
in data with low SNR.

% --- Present main figure, discuss main age results ---
The resulting best fits are overplotted on the original spectra in
Figure \ref{fig:Pops_data_and_magefit}. In general the agreement with
the data is quite good.
The luminosity-weighted mean ages \Tlight\ of these models are given
in Table \ref{tab:Pops_mage}, while their distribution is shown in the
top panel of Figure \ref{fig:Pops_Tlights}. The \Tlight\ values have a
median of $10^{10}$ yr, with 68\% of the values around the median
lying between $10^{9.39}$ and $10^{10}$ yr.

\placefigure{fig:Pops_Tlights}
\notetoeditor{Figure 3 should fit in one column}

The age estimates produced in this analysis, which are conservative as
lower limits, reinforce the conclusion from the single-age analysis
that the majority (17/23) of our nuclear spectra are consistent with a
stellar population considerably older than $10^{9.5}$ yr.  Exceptions
to this finding include the same sources identified as anomalies in
\S\ref{subsec:Pops_results_sage}, with in addition NGC~3489, NGC~3982,
and NGC~4321.
The impact of considering multiple instead of single templates on our
age estimates is also shown in the middle panel of Figure
\ref{fig:Pops_Tlights}.
Allowing multiple templates has an impact on the estimated age only
for those objects that require contributions from both old and young
templates: the best single template appears to be systematically
younger than the luminosity-weighted mean age of the present
population mix.

\placefigure{fig:Pops_mage_vs_sage}
\notetoeditor{Figure 4 should be on top of the page and occupy all its width}

Figure \ref{fig:Pops_mage_vs_sage} shows four examples of how the
inclusion of multiple single-age templates considerably improves the
description of our data.
The need for multiple templates is mostly dictated by the continuum
shape, in that younger model components yield improvement in matching
the blue end for some objects (e.g., NGC~4321, NGC~4203).  Our models
are, however, also very sensitive to strong absorption features in the
spectra of the highest quality (NGC~3368, NGC~5055).
The light weights for the contributing age components of each model
are listed in Table \ref{tab:Pops_mage_lweights} and shown graphically
in Figure \ref{fig:Pops_mage_lweights}.

Although the models of Figure \ref{fig:Pops_data_and_magefit} would
seem quite good, the $\chi^2_\nu$ values listed in Table
\ref{tab:Pops_mage} suggest that in some cases the models are not
altogether satisfactory. Several factors contribute to this situation.

\placefigure{fig:Pops_mage_lweights}
\notetoeditor{Figure 5 should fit in one column}

The most common disagreement with the observations appears around the
Mg~$b$ 5175~\AA\ feature. The quality of the fit in this region
appears to be correlated with the success in fitting the CN-band
absorption around 4200~\AA; NGC~4800 is illustrative of a successful
fit for the two features, while NGC~4459 shows noticeable deviations
in both.
This pattern suggests an $\alpha$-enhancement in the chemical content
of the nuclear stellar populations of many of these sources, and
highlights the need for more sophisticated models accounting for
non-solar abundance ratios.
While the Mg~$b$ and CN features are some of the more obvious regions
showing disagreement, they do not dominate the fit outcome.  Exclusion
of these regions alone from the fit does not substantially decrease
the $\chi^2_\nu$ values (the median value moves from 1.58 to 1.38) and
also does not affect the derived population mixtures. Other weaker
features are also a factor in the elevated $\chi^2$ values.

The other most common source of deviation is found at the bluest ends
of many of our spectra, where our models tend to be brighter than the
data.
In this regard we note two kinds of behavior. For some objects the
shape of the blue continuum of our model appears approximately correct
(e.g., NGC~4459, NGC~4477) while for others there is greater contrast
and a divergence of the model and data at the blue end (e.g.,
NGC~4203, NGC 4450, and to a lesser degree in NGC~3982, NGC~4143).
For most objects in the first class, the SED is already matched by the
oldest, and reddest, template in our library, which, however, is still
too blue in comparison with the data.  Adding more dust extinction
does not help, as it makes the model too red at the long-wavelength
end.
Another possibility is that a component of stars older than 10 Gyr is
present.  As an indicator of how this could change the spectrum, we
note that for the same flux around 5700~\AA, a 20-Gyr-old B\&C
template is roughly 20\% fainter around 3000~\AA\ than the 10-Gyr-old
template.
For the objects in the second class, on the other hand, a combination
of old and young stars does a much better job than a single-age model
in reproducing the overall continuum shape (e.g., see NGC~4203,
Fig. \ref{fig:Pops_mage_vs_sage}); however, the required amount of
$10^{6.5-7.5}$-yr-old stars still does not account for the blue
continuum slope. 
For these objects, we suspect that a featureless continuum may
contribute to their SEDs.

Finally, the STIS error arrays associated with the spectra may
sometimes be underestimating the real uncertainties. 
This seems to be the case in particular for NGC~5055, where inspection
of the model reveals an apparently quite good fit to the data
(Fig. \ref{fig:Pops_mage_vs_sage}), yet $\chi_\nu^2 = 10.45$; the
formal errors seem to be smaller than the intrinsic scatter in the
data.
This situation does not affect our results, in that a global
scale-factor offset in the error arrays does not influence what
population mix provides the best fit; the relative weighting is
unchanged.  For each object, the error spectrum properly gives the
lowest weight to the bluest part of the spectrum, where the data
are noisiest.

\subsubsection{Significance of Multiple Components}

To quantify in more detail whether multi-age templates are really
needed to describe these data, we considered the best-matching models
using one template or using multiple templates as two particular
configurations of the same model, with the single-template models
being a nested realization where 10 of the light weights are set to
zero.
In this situation our two representations of the data would be
consistent with each other if their $\chi^2$ values would differ by
less than the $+3\sigma$ confidence limit corresponding to a $\chi^2$
distribution with a number of degrees of freedom equal to the number
of free parameters in our models \citep[see Theorem C of
\S15.6 of][]{Pre86}. If the $\chi^2$ values differ by less 
than this amount, there is no need for multiple templates.
Our models include 14 parameters, of which 2 describe the kinematics
($V, \sigma$), 1 describes the intrinsic reddening ($A_V$), and 11 are
relative weights of the empirical templates. Although the problem is
not linear in $V$, $\sigma$, and $A_V$, we assume that the number of
data points is large enough for this approach to still be valid.
In order to be as conservative as possible during this evaluation, we
rescaled all $\chi^2$ values assuming that our best multi-age models
have $\chi^2$ equal to the number of useful data points in our nuclear
spectra minus the number of free parameters.
In most of the present cases this procedure is equivalent to stating
that the adopted errors are actually underestimates of the real
uncertainties, which appear justified in light of the previous
considerations.

Among the nuclei requiring multiple templates in Table
\ref{tab:Pops_mage_lweights}, the difference between the $\chi^2$
values for the best single-age and multiple-age models suggests a real
need for a template mix in all of them except for NGC~4245 and
NGC~4800. For NGC~4435, a closer comparison between the model
differences and the error arrays suggests that the evidence for
multiple components is also marginal.
This analysis demonstrates in general that we are not sensitive to
detection of distinct components with light fractions less than a few
percent (see Table \ref{tab:Pops_mage_lweights}), depending on the
quality of the spectra.

We can use these results to gauge the significance of the light and
mass fractions of stars younger than 1~Gyr yr for our sample, which
are also listed in Table \ref{tab:Pops_mage}.
NGC~278 is confirmed to be the youngest nucleus in our sample, and is
best explained with a single-age, $10^{8.5}$-yr-old template.
Excluding the previously mentioned weak and marginal cases, the fits
improve significantly for 9 out of 23 objects when contributions from
stars younger than 1~Gyr are included, with flux fractions larger than
10\% in 8 cases.
For NGC~3489, the fraction of stars younger than 1~Gyr is less
than 10\%, but the large weight for the 1 Gyr template (Table
\ref{tab:Pops_mage_lweights}) and the coarseness of our age grid leave
open the possibility that we may be underestimating the fraction of
stars younger than 1~Gyr.
On the other hand, for NGC~3982, NGC~4143, NGC~4203, and NGC~4450, the
poor match of the blue-continuum shape after introducing young stars
suggests that their presence may be overestimated.
Instead, the presence of AGN continuum would be consistent with the
signatures of broad lines (NGC~4143, NGC~4203, NGC~4450) or strong,
high-excitation emission lines (NGC~3982) found in these objects. All
of these nuclei also show very concentrated light profiles, reminiscent
of unresolved central components (\S4.4).

A better argument to rule out the presence of young stars in
favor of a featureless continuum in these four objects would be to
find evidence for dilution in the absorption-line features.
In order to make our method more sensitive to the absorption
features rather than to the continuum shape, we repeated our analysis
including an additive, low-order polynomial component while solving 
for the template weights (step 2, \S\ref{sec:Pops_method}), and 
suppressing the reddening correction.
The polynomial introduces a considerably greater degree of flexibility
for optimizing the overall continuum shape of the model, but its
presence will tend to dilute the strength of narrow absorption lines,
making it particularly suitable for revealing the presence of an
AGN-like continuum component. 

After refitting all spectra with this prescription, the main results
of our analysis (i.e., light-weight distributions, luminosity-weighted
mean age, and light-fraction of stars younger than 1 Gyr) turned out
to be very similar to those obtained from our previous, best-reddened
models. The only exceptions to this statement are exactly the same
four targets where we already suspected that the presence of a
featureless continuum is more likely than that of $\sim$1-Myr-old
stars.
With inclusion of the polynomial, only a very old population is needed
to describe these nuclei, while the presence of young stars persists
in all the other objects where the reddened models required them.
This outcome suggests that for the latter spectra, and the others
showing only old stellar populations, the polynomial adjustment simply
acts as a substitute for interstellar reddening without affecting
significantly the absorption-line strengths, while in NGC~3982,
NGC~4143, NGC~4203, and NGC~4450 it is actually required by the data
to mimic a blue featureless continuum and to dilute the
absorption-line features.

Yet, it may be argued that allowing the youngest stellar component to
be subjected to a larger amount of dust extinction than the rest of
the stellar population, so as to mimic a dusty nuclear starburst, will
also provide the needed correction for the continuum slope and
dilution of the absorption features.
We explored this possibility in the specific case of NGC~4203. We have
modified our standard multi-component fit to produce two sets of
models, one that allows for a different reddening of the youngest
templates ($< 100$ Myr), and another that includes a featureless
continuum represented by a power law with a slope that is free to vary
and that is also subject to a different reddening. No polynomial
components were used.

Fig. \ref{fig:Pops_Young_vs_PowLaw} shows the result of this
experiment. Both heavily reddened O-stars and an AGN component can
indeed offer almost identical continuum-shape corrections, which
improve the match to the blue end of the spectrum.
Yet the young templates are not entirely featureless and the model
with an AGN component can better explain the data in the spectral
region before and after the Balmer break (between 3600~\AA\ and
3850~\AA). Although the presence of high-order Balmer emission lines
filling the corresponding absorption features could mitigate this
difference, the mismatch in the spectral region on the blue side of
the [\ion{O}{2}] $\lambda$3727 emission line remains larger for the
model with young stars.
Even if  the difference between these  two models is  significant in a
statistical sense, we consider that data with higher SNR are needed to
convincingly separate the  contribution of young stars from  that of a
featureless continuum in this  wavelength range; for the objects where
we  suspect  the presence  of  an  AGN-like  continuum, the  strongest
evidence for this component is  therefore indirect, based on the other
indications of an accretion source as noted above.
We note that Gonz\'{a}lez-Delgado et al. (2004) similarly concluded that
the presence of young stars is unlikely in these nuclei, and pointed
out that using UV spectra is the best way to further investigate this
issue (e.g., Heckman et al. 1997).

\placefigure{fig:Pops_Young_vs_PowLaw}
\notetoeditor{Figure 6 should be on top of the page and occupy all its width}

\subsubsection{Other Considerations}

In \ref{subsec:Pops_obs_bulgecontamin} we estimated that on average
$\sim40$\% of the light in our spectra could originate from bulge
stars, most likely very old ones (e.g., Peletier et al. 1999). If we
assume that such contamination enters only in the fraction of light
from 10-Gyr-old stars that we detect, than its impact on the light
weights of Table 4 can easily be estimated by re-normalizing all light
weights after decreasing the fraction of 10-Gyr-old templates by 0.4,
or set them to zero when this is less than 40\%. Although this would
increase by $\sim70$\% all other light weights, the fractions of light from
stars younger than 1 Gyr would change very little, increasing on
average by only 0.05. The impact on the mass fraction would be even
smaller.

We conclude this section by considering the amount of intrinsic dust
reddening required by our models. The extinction values listed in
Table \ref{tab:Pops_mage} have an average $\overline{A}_V = 0.37$ mag
and a standard deviation of $0.25$ mag.
This result is smaller than the typical $A_V=0.6\!-\!1.0$ mag found by
\citet{Pel99} through analysis of $BIH$ colors measured from {\em HST}
images of disk galaxy bulges.  The origin of this discrepancy is not
clear.
Although the wavelength coverage of our spectra is much less than that
of the broad-band observations of Peletier et al., we consider our
$A_V$ estimates to be fairly robust.\footnote{ For our adopted $A_V$
grid, models with $A_V$ differing from the best-fit value invariably
result in much worse $\chi^2$; consequently the grid step of 0.1 can
be considered a conservative estimate of the uncertainty in $A_V$.  }
The information provided by detailed spectral features in addition to
the broader shape of the blue continuum means that our analysis is
much less susceptible to uncertainties in the stellar population,
while retaining considerable sensitivity to reddening.

One aspect of the bulge stellar populations that we have not yet 
addressed is the effect of different metallicity.  Studies of
abundance gradients in galaxy disks suggest that the regions we are
studying may have very high metallicities \citep[e.g.,][and references
therein]{Vila92}.  This issue is important in the context of
determining $A_V$ values, since super-solar metallicities result in
redder intrinsic templates (see Fig. \ref{fig:Pops_BC03}). A
possibility is that the galaxy spectra are systematically redder than
the solar metallicity templates in part because they have
metallicities $Z > Z_\odot$, and not simply due to interstellar
absorption. Use of nonsolar templates in the fitting process can also
be expected to affect our estimates of \Tlight\ since age and
metallicity are well known to be highly degenerate in their
spectroscopic effects. We explore these issues in the following
section.

\subsection{Super-Solar Metallicity Models}
\label{subsec:Pops_results_mage_ss}

In order to investigate the importance of enhanced metallicity in our
sample nuclei, we repeated our multi-component analysis, including
interstellar reddening, with B\&C templates for super-solar
metallicities, which are limited to $Z = 2.5 Z_\odot$.

Table \ref{tab:Pops_mage_ss} lists these results; the corresponding
\Tlight\ distribution is characterized by a median age of $10^{9.89}$ yr,
with 16\% and 84\% percentiles of $10^{9.23}$ and $10^{10}$ yr, respectively.
These values differ only slightly from the findings reported in
\S\ref{subsec:Pops_results_mage}.
The bottom panel of Figure \ref{fig:Pops_Tlights} further shows that
for all but three objects, the \Tlight\ values shifted by less than
30\% toward younger ages. In fact, the \Tlight\ values of Table
\ref{tab:Pops_mage_ss} are on average only 10\% younger than those of
Table \ref{tab:Pops_mage}.

\placefigure{fig:Pops_magess_vs_mage}
\notetoeditor{Figure 7 should be on top of the page and occupy all its width}

A comparison between the $\chi_\nu^2$ values listed in these two
tables shows that for all but two nuclei, the use of templates with
super-solar metallicities always leads to a better description of the
data, although the improvement is not always statistically
significant.
Figure \ref{fig:Pops_magess_vs_mage} shows two examples of fits where
the improvement resulting from use of the 2.5$Z_\odot$ templates is
notably strong (NGC~4459, NGC~4477), along with two examples where the
improvement is more modest (NGC~4800, NGC~278). The case of NGC~4459
is particularly instructive in showing details that are better fit
with the high-metallicity templates.  The Mg~$b$ 5175~\AA\ feature
benefits from using these models; likewise, the whole region between
4400~\AA\ and 4800~\AA\ is better reproduced with the super-solar
templates, in particular thanks to stronger Fe 4531~\AA\ and Ca
4668~\AA\ absorption lines.
The high-metallicity templates are not sufficient to explain the
strength of the CN band around 4200~\AA, however, consistent with the
idea that this mismatch highlights the need for non-solar
abundance ratios.
Likewise, the 2.5$Z_\odot$ templates cannot account for the
discrepancies in the bluest regions of our spectra noted in
\S\ref{subsec:Pops_results_mage}, suggesting again that the oldest
nuclei in our sample may have stellar populations older than 10 Gyr,
or that in cases like NGC~4203 and NGC~4450 the presence of young
stars may still be overestimated.

In Table \ref{tab:Pops_mage_ss} we list the summed light and mass
fraction of stars younger than 1 Gyr for the super-solar models
affording the best fits.  Use of the high-metallicity templates does
not change our conclusions concerning which nuclei show significant
($\ga 10\%$) evidence for a young stellar component.
The table also lists values for $A_V$, which, as expected, are lower
than the $Z_\odot$ case, with $\overline{A}_V = 0.18 \pm 0.04$ mag.  A
suspicious aspect of the $A_V$ distribution is the fact that 10 out of
23 nuclei are best fit with $A_V = 0$ mag. This finding suggests that
the $Z = 2.5 Z_\odot$ templates are actually {\em redder} than our
spectra in many cases, which might imply that a metallicity of
2.5$Z_\odot$ is too high.  The fact that the detailed absorption
features are better fit by the 2.5$Z_\odot$ models while the overall
continuum shape is apparently better represented by the $Z_\odot$
models suggests that we are encountering a fundamental limit in the
accuracy to which the B\&C models can be applied to the current
problem.  The results nonetheless allow us to conclude that typical
metallicities fall within the interval of (1--2.5)$Z_\odot$ for the
inner bulges of our sample galaxies.

\placefigure{fig:Pops_upplim}
\notetoeditor{Figure 8 should be on top of the page and occupy all its width}

\subsection{Contribution from Nuclear Power-Law Sources 
or Young Star Clusters}

\label{subsec:Pops_results_mage_upplim}

Our multi-age fits indicate that a significant fraction of the nuclei
in our sample have a component of very young (few-Myr old) stars, or
alternatively, a featureless blue continuum component as may arise
from an accretion source.
In this section we present quantitative bounds on the contributions of
such components along with their significance, evaluated by a
$\Delta\chi^2$ analysis as in our study of multiple-age components
(\S\ref{subsec:Pops_results_mage}).
We adopted as our fiducial description of the nuclear spectra the
multi-age models with solar metallicity derived in
\S\ref{subsec:Pops_results_mage}.

For each object we obtained fits by fixing the fraction of total blue
light $f_{blue}$ contributed by a $10^6$-yr-old population, and
repeating the fit with different $f_{blue}$ values to produce a model
sequence. For a given fit, the relative contributions of the other age
components and the common extinction value were left as free
parameters. We then compared the results with the best fits, and
derived $\pm 3\sigma$ confidence limits for $f_{blue}$ as in \S 4.2,
where in this case the $\chi^2$ values follow a $\chi^2$ distribution
for one degree of freedom.
The same approach was used to determine the best-fit contribution and
3$\sigma$ bounds for a featureless continuum, which we represent by a
power law with $f_{\nu} \propto \nu^{-\alpha}$ or $f_{\lambda} \propto
\lambda^{\alpha -2}$. 
Our limited spectral baseline makes it difficult to independently
constrain $\alpha$, and consequently we adopted a fixed value of
$\alpha = 0.5$, which is representative of power-law slopes observed
in luminous AGNs \citep[e.g.,][]{Zhe93}.
The fraction $f_{blue}$ is calculated for the bandpass 3050--3200~\AA,
where a component of young stars or an AGN continuum is likely to be
most prominent.
In order to derive conservative bounds, we rescaled all $\chi^2$
values as done is \S\ref{subsec:Pops_results_mage}, which has the
result of widening the calculated confidence limits.

The results, shown in Figure \ref{fig:Pops_upplim}, indicate that the
flux contribution from a ``blue component'' is very small ($\la10\%$)
in roughly half of our nuclear spectra, regardless of the stellar or
AGN nature.

A number of other nuclei (e.g., NGC~3982, NGC~4143, NGC~4203, and
NGC~4450; see also \S\ref{subsec:Pops_results_mage}) clearly require a
non-zero AGN-like continuum (Fig. \ref{fig:Pops_upplim}{\it a}).
The presence of probable AGN continuum emission is in accord with the
signatures of broad lines (NGC~4143, NGC~4203, NGC~4450) or strong,
high-excitation emission lines (NGC~3982) found in the same objects.
The fits also suggest that a young-star component is not strictly
needed in NGC~3982, NGC~4143, NGC~4203, and NGC~4450.
With such an AGN component these nuclei require only a 10 Gyr stellar
population.
In other objects an AGN-like continuum is perhaps needed, but the SNR
is too low to be conclusive (NGC~3351, NGC 4321).

Excluding the probable AGNs (shown with dotted lines in Figure
\ref{fig:Pops_upplim}{\it b}), the 3$\sigma$ maximum contribution from
a $10^{6}$-yr-old population has a median value of only 6\% for the
sample, which represents the sensitivity limit of our experiment for
detecting such young stars. For comparison, in the case of our Milky
Way we would not have been able to detect the light emitted by its
young central clusters, as these contribute only $\sim 1$\% of the
light at $\sim 3000$~\AA\ coming from the central 10 pc (D. Figer,
2003, private communication).

The STIS acquisition images for our target galaxies reveal that the
objects requiring a featureless component all show a high degree of
concentration in their nuclear surface brightness distribution.
In the case of NGC~4203, this correspondence supports the exclusion of
its central unresolved light component from the stellar budget while
measuring its central SMBH mass \citep{Shi00,Sar01}.
No such signature is apparent from inspection of images for sources
with spectral fits implicating a Myr-old stellar population, with the
exception of the youngest object in our sample, NGC~278, which
displays a nuclear cluster like the ones discovered by \citet{Caro98}
and \citet{Boe02}.
The work by Carollo \ea indicates that compact nuclear components can
often be rather subtle in their photometric signature, and detection
of these structures in images such as we have available can be
seriously complicated by dust absorption. It is therefore possible
that compact clusters are present in a large fraction of our sample
galaxies; near-infrared images would be very helpful in addressing
this issue.

We conclude this section by asking whether the upper limits on young
stellar populations would be accompanied by emission of ionizing
photons sufficient to explain the observed emission-line fluxes,
collected either within the same {\it HST\/}-STIS aperture or by
ground-based observations.

We used the 1-Myr model SEDs along with the derived upper limits on
their blue light contribution to estimate the maximum production rate
of ionizing photons in each source, in order to predict the H$\alpha$
flux, by assuming that all ionizing photons are absorbed and that Case
B recombination applies \citep{Hum87}. This derivation includes the
extinction correction for the blue continuum that we obtained from our
fit, and assumes that a negligible fraction of the ionizing photons
are intercepted by dust.  The resulting predictions are thus
conservative in estimating the {\em maximum} H$\alpha$ fluxes that
could plausibly result from \ion{H}{2} regions powered by young
stars. We finally applied to these hypothetical fluxes the foreground
Galactic extinction along each object direction (Table
\ref{tab:Pops_GalSample}).
Table \ref{tab:Pops_ionizingfluxes} lists the theoretical predictions
from this exercise along with the observed H$\alpha$ fluxes observed
in our G750M spectra. The H$\alpha$ measurements were obtained after
subtraction of the stellar continuum as part of a larger study of the
nebular properties of the sample nuclei presented by
\citet{Shi04b}. The listed H$\alpha$ fluxes have not been corrected for
extinction internal to the source galaxy. Reddening estimates from the
measured H$\alpha$/H$\beta$ ratio and recombination theory are
available for fewer than half of our target nuclei because of
limitations in the data.\footnote{Studies of other star-forming
galaxies indicate that reddening of the optical nebular emission is
generally comparable to or somewhat greater than that of the
associated blue starlight \citep[e.g.,][]{Cal97}, and this behavior is
weakly confirmed in our data.} By neglecting internal extinction of
H$\alpha$, we are again conservative in gauging the extent to which
something other than stars is required to power the nebular emission.
A comparison of the resulting predicted versus observed H$\alpha$
fluxes is shown graphically in Figure \ref{fig:Pops_Ha_predvsobs},
where the emission-line classification derived from ground-based
observations \citep{Ho97} is indicated by the symbol type.

\placefigure{fig:Pops_Ha_predvsobs}
\notetoeditor{Figure 9 should fit in one column}

It is important to keep in mind that \ion{H}{2} regions can be
subject to large amounts of extinction, up to several magnitudes
\citep[e.g.,][]{Rod99}.  More O-stars could thus be present in these
nuclei without being detected, if they are substantially more obscured
than the rest of the stellar population.  However, such deeply
embedded stars would not contribute significantly to the ionizing flux
powering the observed nebulosity unless the geometry were fine-tuned
so that the stars were obscured only along our line of sight.

Figure \ref{fig:Pops_Ha_predvsobs} shows results that are generally
consistent with our expectations for the role of accretion power, as
indicated by the emission-line classification of the nuclei.  For
three of the four Seyfert nuclei in our sample, the observed H$\alpha$
flux exceeds the predicted contribution from young stars, and the same
statement probably applies to the fourth object, NGC~3982, since as
noted earlier the young stellar component is likely overestimated in
this source.
Two of the LINER 1 nuclei (NGC~4203, NGC~4450) are formally consistent
with ionization by stars, although the allowed flux from young stars
is again probably overestimated; consistency with stellar
photoionization is at best marginal for the remaining two LINER 1s
(NGC~2787, NGC~4143) under the current optimistic assumptions.  A
majority of the LINER 2s also show only marginal consistency with
starlight as the sole source of ionization.
The situation is mixed for the transition sources: two are consistent
with stellar photoionization, two are probably not, and a further two
are indeterminate because of upper limits in both coordinates.
The \ion{H}{2} nuclei that are detected in H$\alpha$ are consistent
with ionization by hot stars.

Considering that the emission-line fluxes obtained in ground-based
measurements \citep{Ho97} over apertures an order of magnitude larger
in radius contain 10--100 times as much emission as measured here
\citep[see][]{Shi04a}, none of these galactic nuclei contain enough
very young stars in their central $\sim10$ pc to explain the nebular
luminosity observed over $\sim 100$ pc scales.
The upper limits for the power law shown in Figure 7{\em b}, however,
translate into ionizing photon production rates that exceed the
large-scale nebular requirements by a factor of 100 or more, when a
simple extrapolation of the continuum is assumed.  The real production
rate will be very sensitive to the details of the continuum shape, but
in general we can conclude that an accretion-powered continuum can in
principle account for the nebular luminosity in these nuclei.

%--- DISCUSSIONS AND CONCLUSIONS
\section{Discussion}

The analysis described above indicates that the majority of stars
within the central few tens of pc in our sample galaxies are many Gyr
old. These results provide some new insights into the histories of
galaxy nuclei. If our limited sample is representative of galaxy
bulges, the majority of the current stellar mass in the central
parsecs was assembled into stars $\sim 10^{10}$ yr ago, and little
star formation has occurred subsequently. Quantitatively, if we adopt
a relatively high value of 2\% (Table \ref{tab:Pops_mage}) for the
present mass fraction of stars younger than 1 Gyr as representative of
star formation over the intervening time, then at most 20\% of the
nuclear stellar mass could have been built in the last 10 Gyr.
The bulk of the stars thus formed at an epoch similar to, or somewhat
earlier than, that of the ``quasar era'' ($z \approx 2$--3). This
inference is in accord with quasar metallicity studies indicating that
luminous accretion episodes are accompanied or preceded by vigorous
star formation \citep{Ham99}, and hence consistent with a picture
wherein the formation processes of galaxy spheroids and SMBHs are
tightly linked.\footnote{Here and in the remainder of this discussion
we excluded from the objects containing stars younger than 1 Gyr those
nuclei where the formal presence of young stars suggested by our
multiple-component analysis is either overruled by indirect evidence
of a power law (NGC~3982, NGC 4143, NGC 4203, NGC 4450), or just
marginal (NGC~4435).} 

A possible explanation for the lack of substantial younger populations
may be advanced by considering impediments to star formation in close
proximity to a SMBH. The tidal radius is $r_t \approx R_{cloud}
(M_{\rm BH}/M_{cloud})^{1/3} \approx (M_{\rm BH}/\rho_{cloud})^{1/3}$,
and the densities within the inner few-pc molecular regions of our
Galaxy are $n_{\rm H_2}=10^4-10^7\,{\rm cm}^{-3}$ \citep{Mor96}. If
these values are typical of other galaxies, even the densest
components will be subject to strong tidal limitations. For example, a
molecular cloud core with $n_{\rm H_2}=10^7\,{\rm cm}^{-3}$ would be
torn apart if wandering closer than 7.3 pc from a $10^8\,\Msun$ SMBH.

Our spectroscopic findings are in general accord with the results of
{\em HST} imaging surveys by \citet{Pel99} and \citet{Caro01,Caro02}.
Peletier et al. find that bulge nuclei are substantially obscured by
dust and very seldom show blue colors, and although the samples of
Carollo et al. include only spiral galaxies, our results are also
consistent with their interpretation of their color-color diagrams as
showing that most of the nuclei look mildly obscured by dust and are at
least 1~Gyr old.
Our spectroscopic analysis suggests that interstellar reddening may be
less than reported by Peletier et al.

In terms of morphology, \citet{Caro98} find that $\sim 50$\% of spiral
galaxies contain a distinct nuclear component. An interesting question
is whether these structures correspond to sites of recent star
formation.
We cannot directly test this idea due to the limitations of our
imaging data for the present sample. However, we can use the upper
limits on young stars from our spectra to compute corresponding
broad-band magnitude limits, in order to gauge whether we are
sensitive to the types of clusters found in previous work.
A conservative threshold corresponding to detection of a 1-Gyr
component providing 10\% of the total measured starlight for one of
our sample galaxies translates on average to a maximum detectable
absolute $V$ magnitude of $-10$.
Although the distinct nuclei found by Carollo et al. often extend
beyond the 8 pc radius typically subtended by our observations, the
vast majority of the published nuclei are much brighter than $M_V =
-10$ mag, suggesting that we can detect such components if they are
indeed 1 Gyr old.
The fact that the fraction of spiral galaxies with distinct nuclear
components ($\sim 50$\%) is rather similar to that of our spirals
showing strong indications for the presence of stars younger than 1
Gyr (5/14) suggests that these objects could in fact contain young
stellar clusters.
The fraction of nuclei with young stars is lower for our lenticular
galaxies (1/9), although this may reflect a different sensitivity
limit for this galaxy type. For a given cluster luminosity the degree
of contrast between the cluster and its surroundings will be reduced
in early-type galaxies with large bulges; the problem is exacerbated
by the fact that bulges tend to have cuspy surface brightness
profiles.

Star clusters will also tend to be luminous when younger, which would
aid in their identification in imaging.  This phenomenon is manifested
in the case of NGC~278, which is unique in our sample for its
prominent Balmer-dominated spectrum and distinct nuclear component in
surface photometry, both consistent with a recently formed cluster.
A more general phenomenon of recurrent star formation associated with
a nuclear cluster, resulting in variation in the prominence of
spectroscopic signatures of young stars, would be in accord with
recent studies of late-type disk galaxies by \citet{Boe02} and
\citet{Wal04}. Their work indicates that nuclear star clusters are
common in such galaxies, that the clusters frequently exhibit
spectroscopic signatures of young stars, and that the masses and
detailed spectral properties are consistent with multiple episodes of
star formation. The discovery of these clusters and the study of their
stellar population is aided in these systems by the absence of a
significant bulge.

Although signatures of young stars are found with somewhat greater
frequency in spirals compared with lenticulars in our sample, there is
not a strong correlation between the presence or strength of these
components and Hubble type considered alone.  Likewise, there is no
correlation with stellar velocity dispersion or galaxy luminosity.
However, other aspects of our sample reinforce the picture that
contamination by bulge light is an important limiting factor for the
detection of young nuclear clusters. Specifically, we note that the
only S0 galaxy with detected young stars is our closest object, while
the most distant galaxy with young stars belongs to the latest Hubble
type in our sample (Sbc).
As a group, then, the galaxies found to be hosting stars younger than 1 Gyr 
in their nuclei tend to be closer than the other objects in our sample,
with a mean distance of 9.7 Mpc as opposed to 15.2 Mpc.

\placefigure{fig:Pops_with1GyrNuclClass}
\notetoeditor{Figure 10 should fit in one column}

Interestingly, the presence of a component of stars younger than 1 Gyr
appears connected also to the emission-line classification of the
nucleus (Fig. \ref{fig:Pops_with1GyrNuclClass}), where the
classifications are those listed in Table 1; for objects with dual
classifications (NGC~3489, NGC~4435), we adopt the type listed first,
which is marginally preferred.
Evidence of young stars is found in \ion{H}{2}, transition, and LINER
2 nuclei, but not in LINER 1s or Seyferts. As discussed in
\S\ref{subsec:Pops_results_mage} and
\S\ref{subsec:Pops_results_mage_upplim}, some LINER 1s and Seyferts do
show evidence of excess blue light (NGC~3982, NGC~4143, NGC~4203,
NGC~4450), but this component is probably best described by a nonstellar power
law rather than by young stars.  The remaining LINER 1s and Seyferts are best 
fit by a purely old stellar population (NGC~2787, NGC~4477, NGC~4501,
NGC~4698). This result is consistent with a picture in which Seyferts
and the majority of LINERs are accretion-powered systems. The
association of young populations with transition or LINER 2
classifications suggests that stellar phenomena may have some role in
influencing the observed nebular emission, although the results in
Figure \ref{fig:Pops_Ha_predvsobs} imply that energetic processes
other than simply photoionization by O-type stars would have to be
involved. Another noteworthy aspect of Figure
\ref{fig:Pops_with1GyrNuclClass} is the fact that the majority of
\ion{H}{2} nuclei do {\em not} show evidence for young stars in the
{\em HST} aperture. Star formation in these objects apparently occurs
primarily in circumnuclear regions subtended by the larger apertures
($\sim 100$ pc) characteristic of ground-based observations, and this
picture is confirmed by spatial measurements of the nebular emission
\citep{Shi04a}.
Our findings suggest that this could be the case also for Seyfert~2
nuclei, for which evidence of a $\sim$100-Myr-old stellar population
at $\sim$100 pc scales has been reported in a number of cases
\citep[e.g.,][]{Schm99}.
The existence of a connection between the presence of young stars and
the emission-line classification is consistent with the results of
Gonz\'ales-Delgado et al. (2004), who find that the incidence of young
stars in weak-[\ion{O}{1}] low-luminosity AGNs ([\ion{O}{1}]/H$\alpha
\leq 0.25$) is larger than in strong-[\ion{O}{1}] nuclei.

\section{Conclusions}

We have fitted stellar population templates to interpret the spectra
for the inner $\sim 10$ pc of a sample of galaxies with a variety of
Hubble types and nuclear emission classifications.
The primary result of our analysis is that the large majority of
nuclei have spectral energy distributions consistent with simply old
stellar populations. Only one object out of 23 is dominated by a young
cluster of stars less than 1 Gyr old, which is also photometrically
distinct.
When using multi-age population fits to reflect more complex formation
histories, we find a distribution of luminosity-weighted mean ages
across the sample that is strongly peaked around 10 Gyr. This result
persists if we account for dust reddening and super-solar metallicities
as contributors to or possible alternative explanations for the quite
red observed spectra. For most sample galaxies, only negligible
amounts of stars can have formed in the central 10 pc during the
second half of the Universe's life.

There is, however, a significant fraction of galactic nuclei that show
modest amounts of young stars: evidence for stars younger than 1 Gyr
is found in $\sim25$\% of the cases. Very young stars, $\sim1$ Myr,
are not required by the spectral template fits for any object, and
their possible contribution to the flux at the blue end of our spectra
($\sim 3100$~\AA) is limited to less than 6\% on average.
With these upper limits, the production of ionizing photons by very
young stars falls short of the requirements for powering the nebular
emission in the central 10 pc of the Seyfert nuclei and most of the
LINERs in our sample. Furthermore, none of these 23 galactic nuclei
contain enough very young stars to explain the nebular luminosity
observed over $\sim 100$ pc scales, as typically observed from the
ground.
This shortfall of stellar UV photons holds true regardless of internal
dust extinction.
The presence of a young population ($\lesssim$1 Gyr) appears, however,
correlated with the classification of the emission-line ratios.

In Seyferts and most LINERs our findings are naturally explained, if
ionization comes largely from AGN accretion, not stars. For \ion{H}{2}
nuclei, our results imply that much of the central star formation is
actually circumnuclear --- it occurs in the range of a few 10 pc to a
few 100 pc.

We note in closing that even for nearby galaxies at {\it HST} resolution, the
contamination from stars in the central portion of the surrounding
bulge is an important limiting factor. We could not have detected a
quantity of young stars as small as found at the center of the Milky
Way.

\acknowledgements M.S. is grateful to Don Figer, Harald Kuntschner,
John Magorrian, Dan McIntosh, Reynier Peletier, Mark Whittle, and
Sukyoung Yi for their suggestions. This work was supported financially
through NASA grants GO-07361 and GO-09788 from the Space Telescope
Science Institute (STScI), which is operated by AURA, Inc., under NASA
contract NAS 5-26555.  Research by A.J.B. is supported by NASA through
Hubble Fellowship grant \#HST-HF-01134.01-A awarded by STScI. The work
of L.C.H. is supported by the Carnegie Institution of Washington.

%%%%%%%%%%%%%%%%%%%%%%%%%%%%%%%%%%%%%%%%%%%%%%%
%
% FIGURES
%
%%%%%%%%%%%%%%%%%%%%%%%%%%%%%%%%%%%%%%%%%%%%%%%

 \setcounter{figure}{0}
 \begin{figure}\plotone{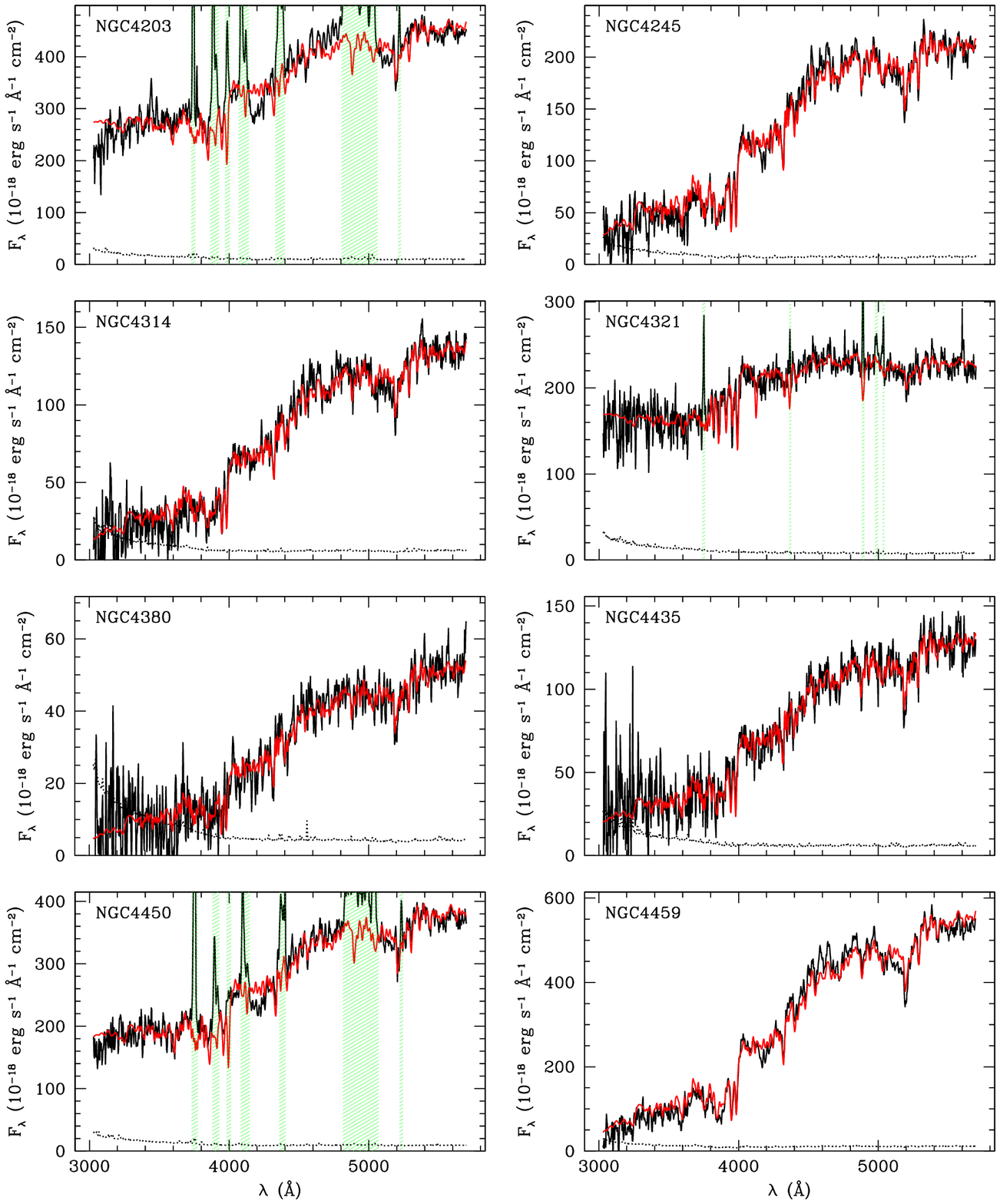}\caption{
 Continue
 }\end{figure}

 \setcounter{figure}{0}
 \begin{figure}\plotone{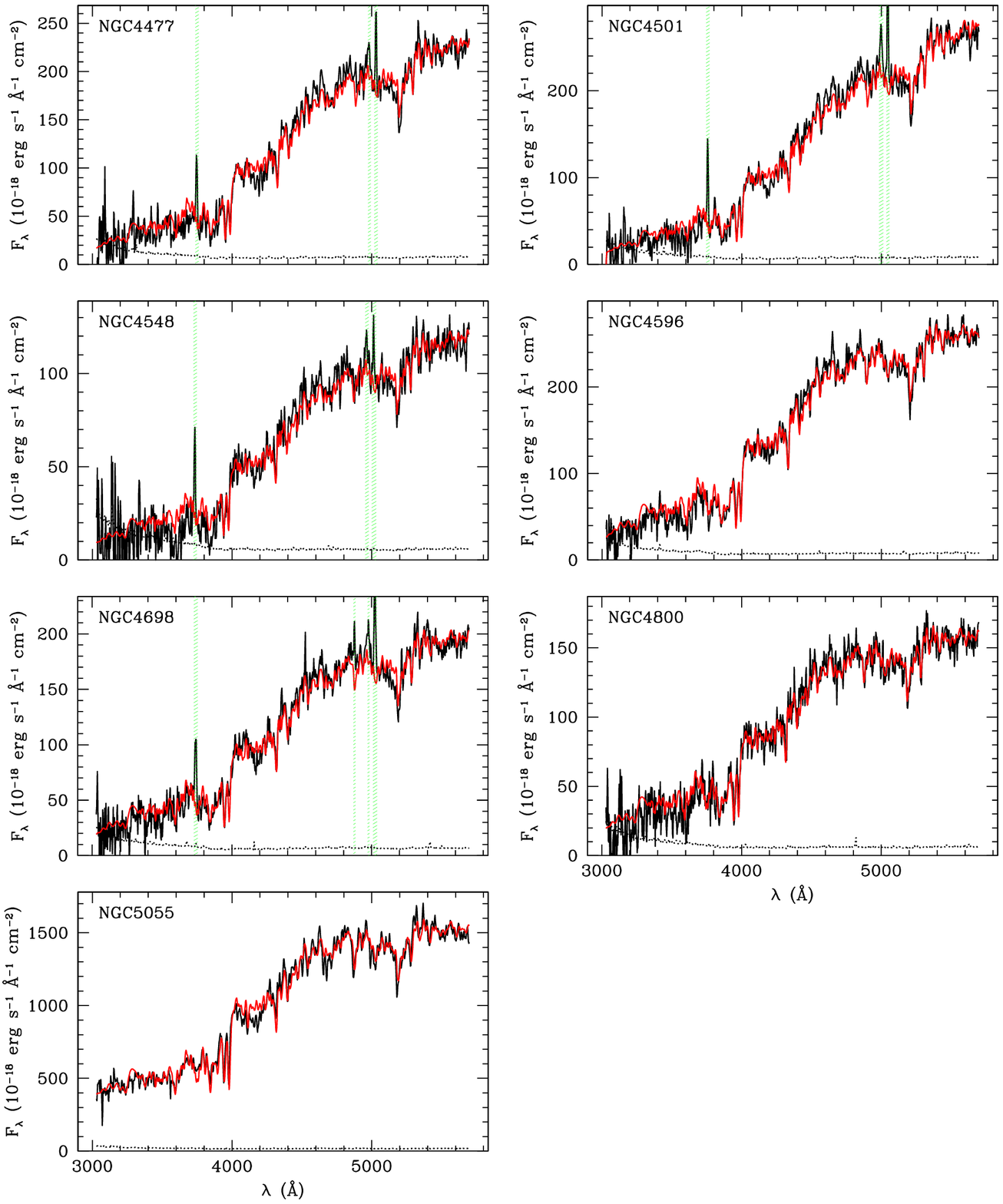}\caption{
 Continue
 }\end{figure}

 \begin{figure}\plotone{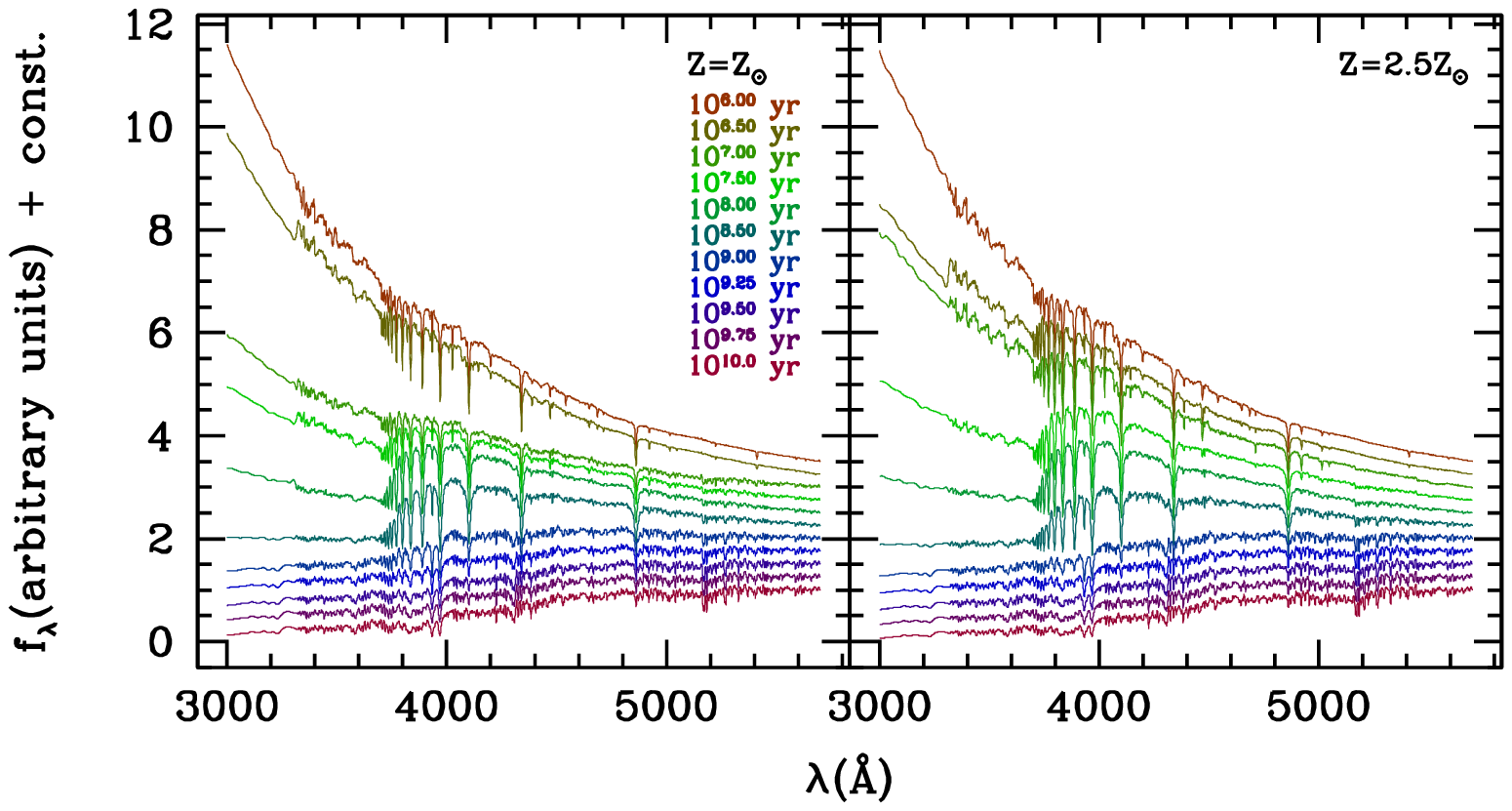}\caption{
 Bruzual \& Charlot (2003) model SEDs for single-starburst stellar
 populations of different ages and solar ({\it left}) or 2.5 times
 solar ({\it right}) metallicities. The models are shown for the
 wavelength range matching our data, and for convenience of display
 they are normalized at 5700~\AA\ and offset by a
 constant.\label{fig:Pops_BC03}
 }\end{figure}

 \begin{figure}\plotone{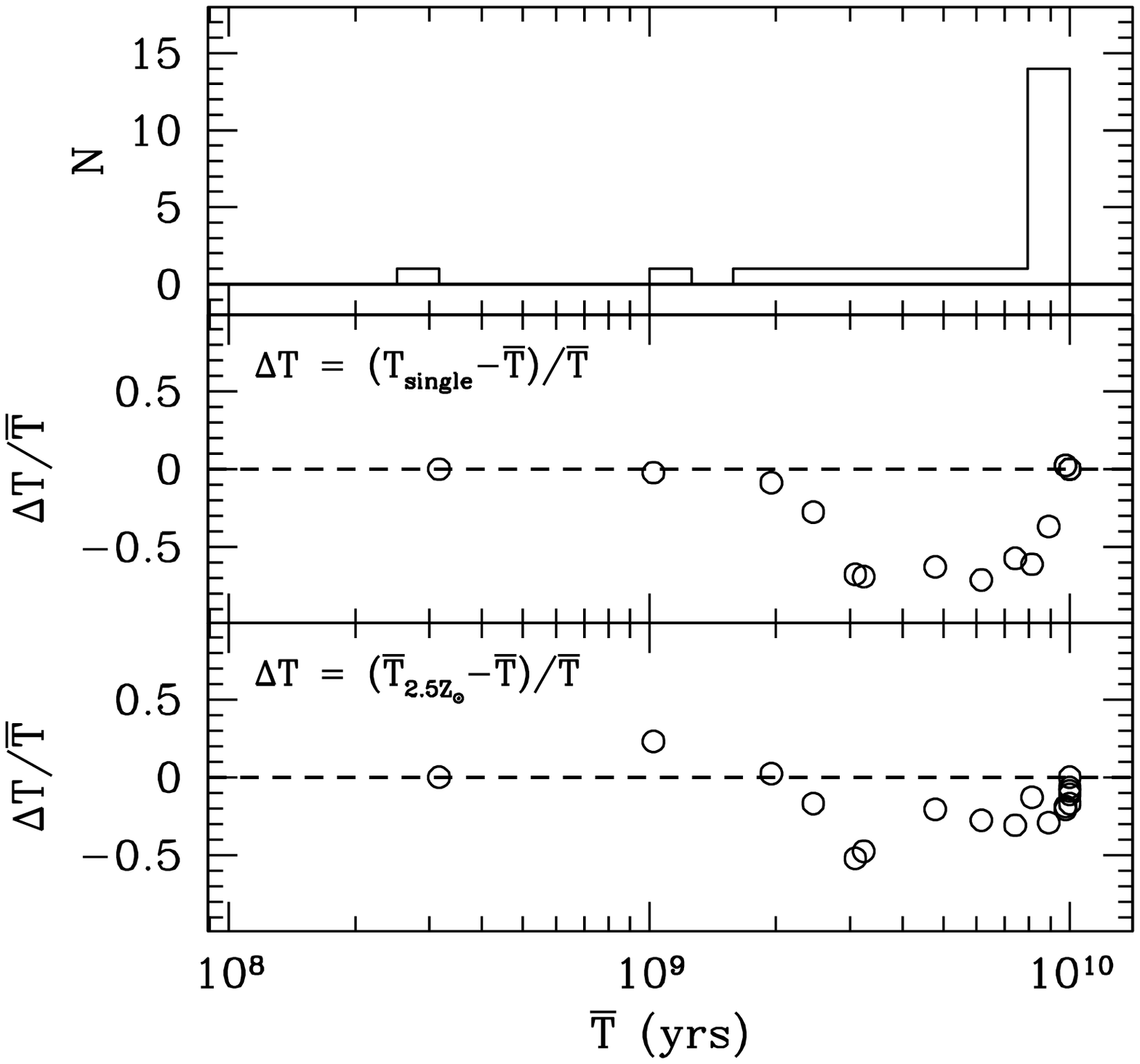}\caption{
 Distribution of the luminosity-weighted age estimates obtained from
 the best-fitting multiple-starburst models ({\it top panel}), which
 are also compared to the age of the best single-age models of \S4.1
 ({\it middle panel}) and the luminosity-weighted ages of \S4.3, from
 multiple-starburst models of super-solar metallicity ({\it bottom
 panel}).\label{fig:Pops_Tlights}
 }\end{figure}

 \begin{figure}\plotone{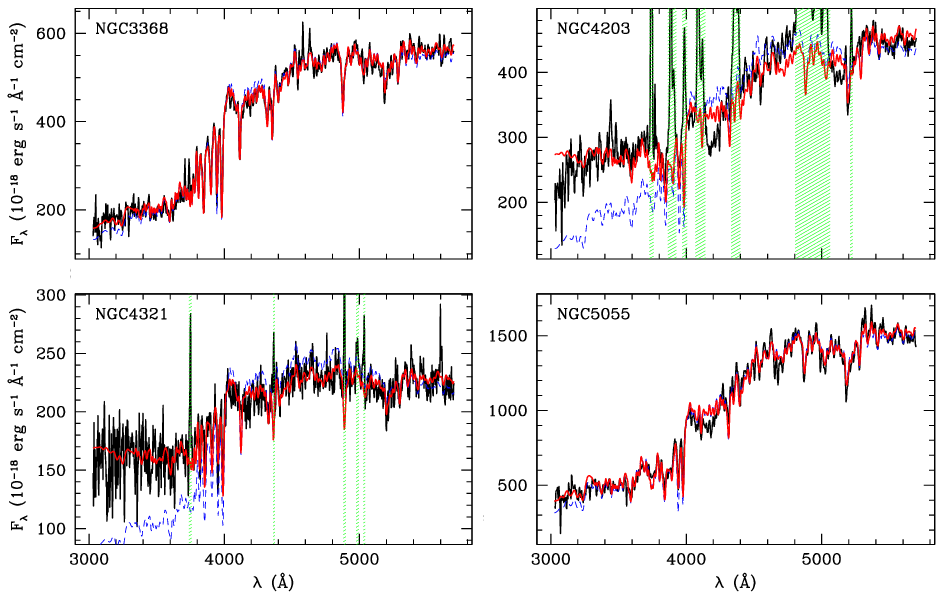}\caption{
 Examples of fits obtained with multiple-starburst models ({\it red
 lines}) versus single-age models ({\it blue dashed
 lines}).\label{fig:Pops_mage_vs_sage}
 }\end{figure}

 \begin{figure}\plotone{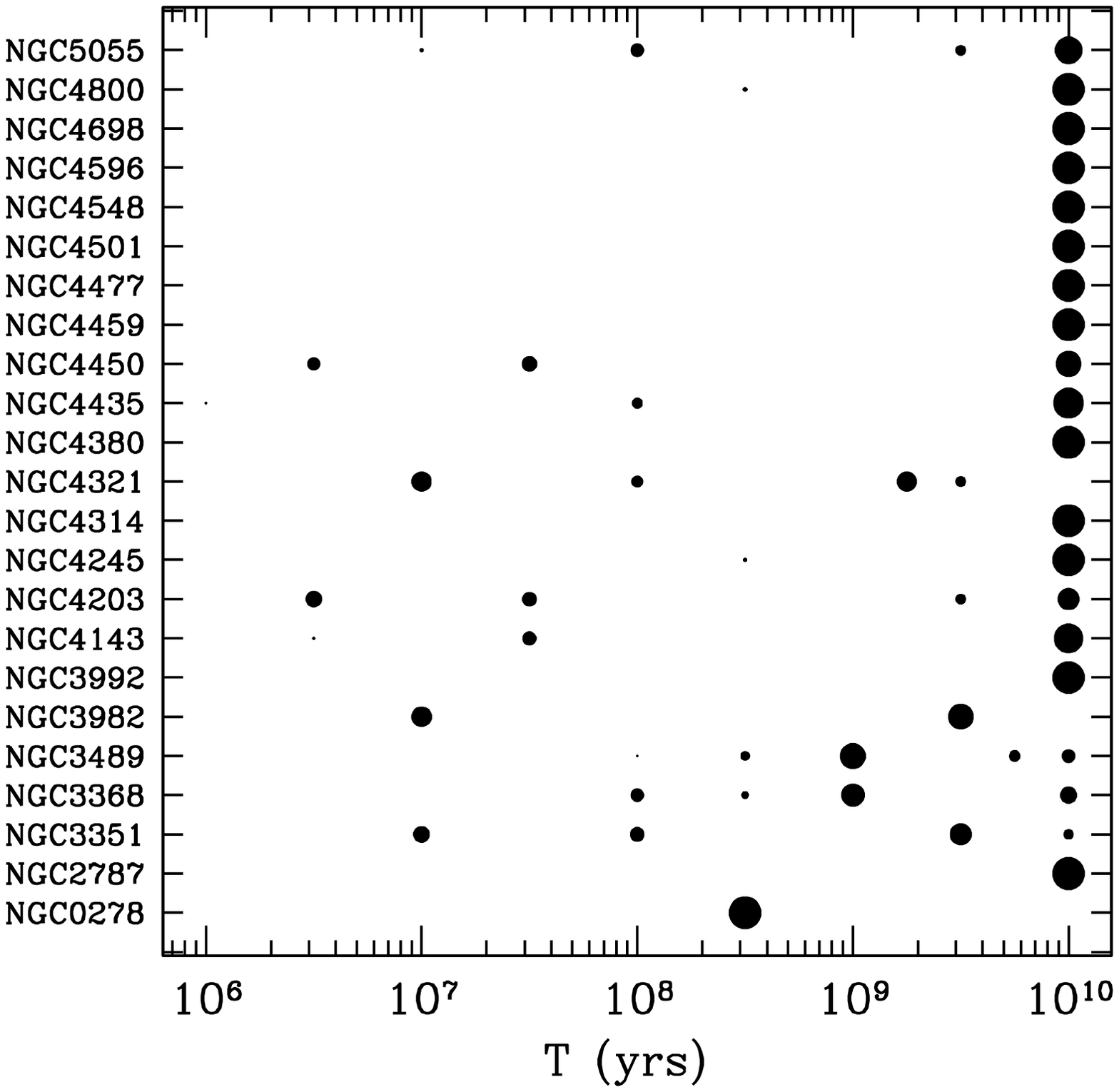}\caption{
 Graphical visualization of the relative light contribution of each
 single-age model to the best multiple-starburst models. The area of
 the {\it filled dots} is directly proportional to the relative
 weights listed in Table\ref{tab:Pops_mage_lweights}.\label{fig:Pops_mage_lweights}
 }\end{figure}

 \begin{figure}\plotone{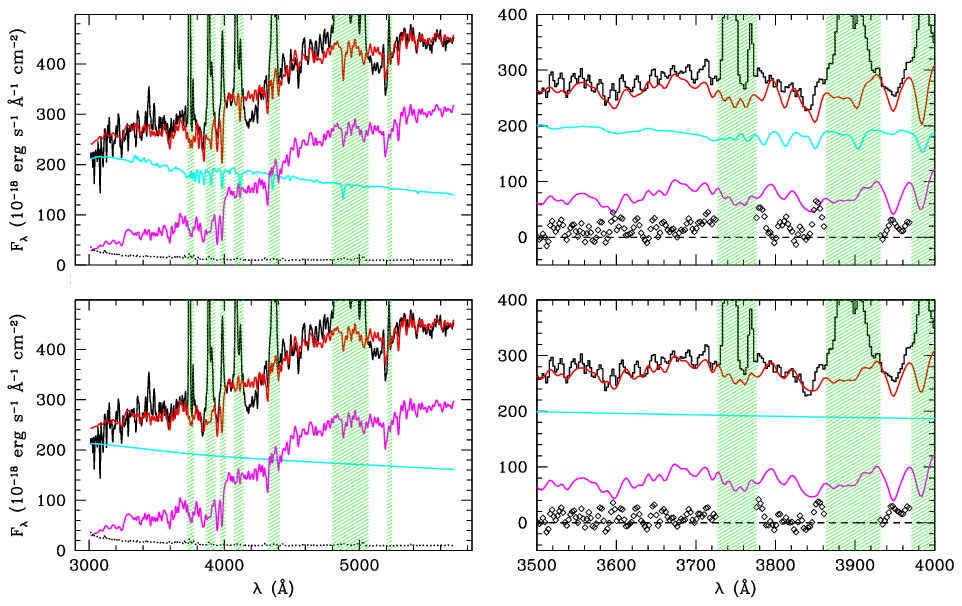}\caption{
 Young stars vs. AGN continuum in NGC~4203. {\it Left panels\/}: As
 for Fig. 1 but now showing separately also the contribution of
 intermediate-old populations ({\it purple lines}) and of either young
 stars ($< 100$ Myr, {\it top panels, light-blue lines\/}) or of the
 best power-law continuum ({\it bottom panels, light-blue
 lines\/}). Both young stars and featureless continuum are allowed to
 be subject to a different amount of dust extinction than the rest of
 the stellar population. {\it Right panels\/}: As in the left panels,
 but now showing the fits in the wavelength region on either side of
 the Balmer break, with the corresponding residuals ({\it
 diamonds}). The model with independently reddened young stars
 requires them to be subjected to a considerable amount of extinction
 ($A_V=2.2$ mag) and to contribute 55\% of the light (from the
 3.2-Myr-old template). The model with an AGN continuum needs only
 10-Gyr-old stars in addition to a power-law with slope $\alpha=1.3$
 reddened by $A_V=0.2 mag$ \label{fig:Pops_Young_vs_PowLaw}
 }\end{figure}.

 \begin{figure}\plotone{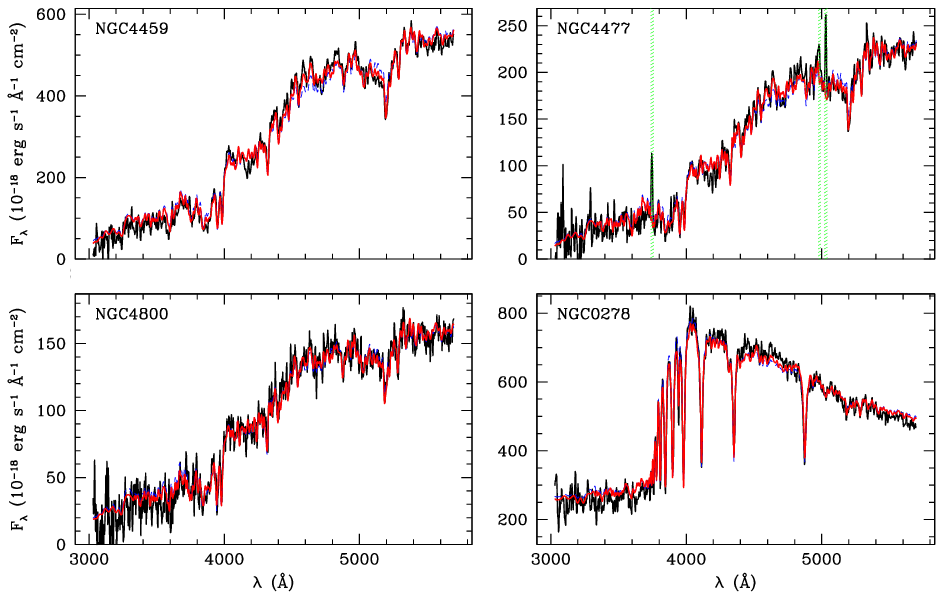}\caption{
 Examples of multiple-starburst models with super-solar
 ($Z=2.5Z_\odot$, {\it red lines}) or solar ($Z=Z_\odot$, {\it blue
 dashed lines}) metallicity.\label{fig:Pops_magess_vs_mage}
 }\end{figure}

 \begin{figure}\plotone{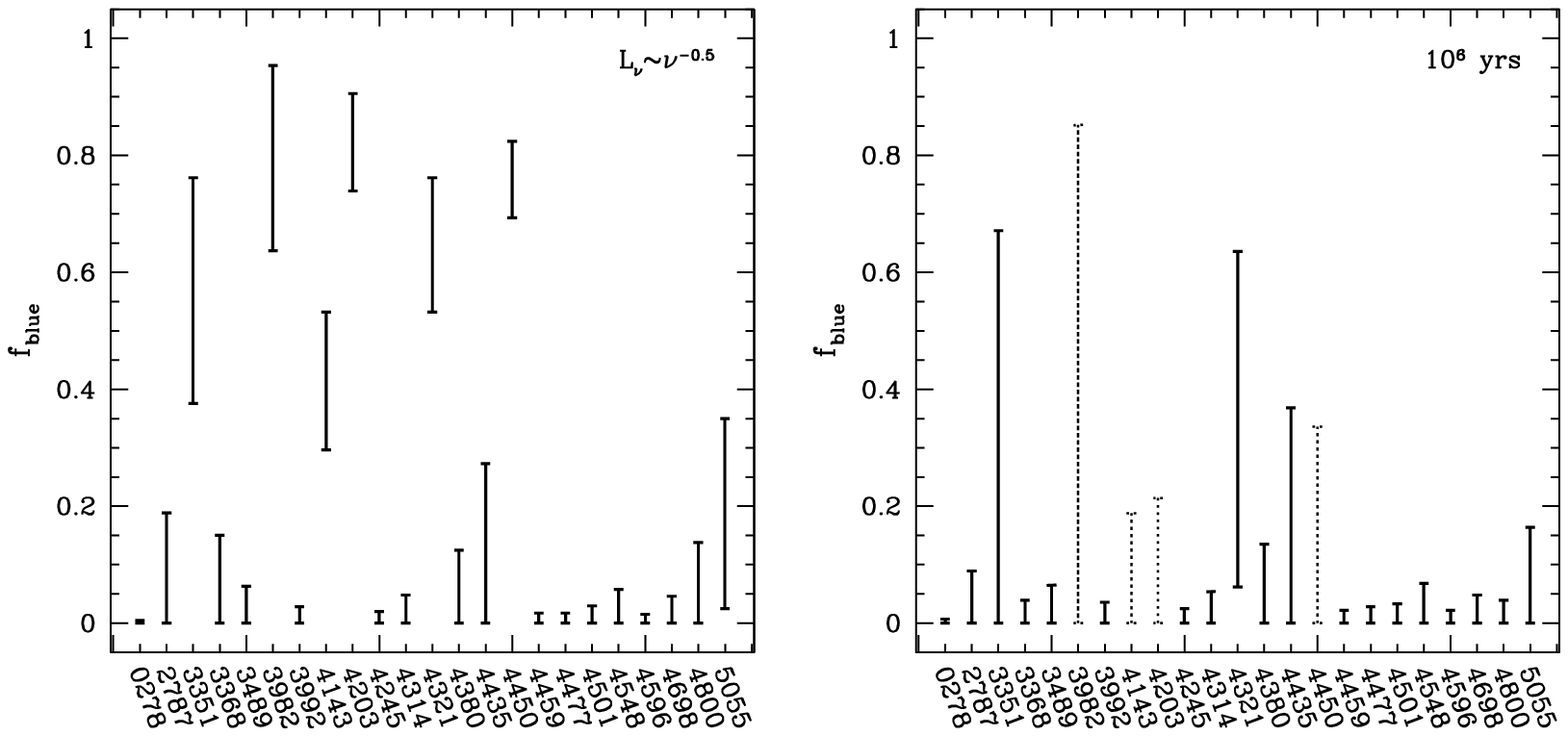}\caption{
 Confidence limits (3$\sigma$) for the flux fraction in the range
 3050--3200~\AA\ from a power-law featureless continuum (for
 $L_{\nu}\propto\nu^{-0.5}$, {\it left panel}), or a single,
 $10^6$-yr-old, starburst population model ({\it right panel}). The
 dotted lines identify cases where the need for young stars is most
 likely overestimated by the models of
 \S\ref{subsec:Pops_results_mage}. Each number is the NGC name of the
 galaxy.\label{fig:Pops_upplim} }\end{figure}

 \begin{figure}\plotone{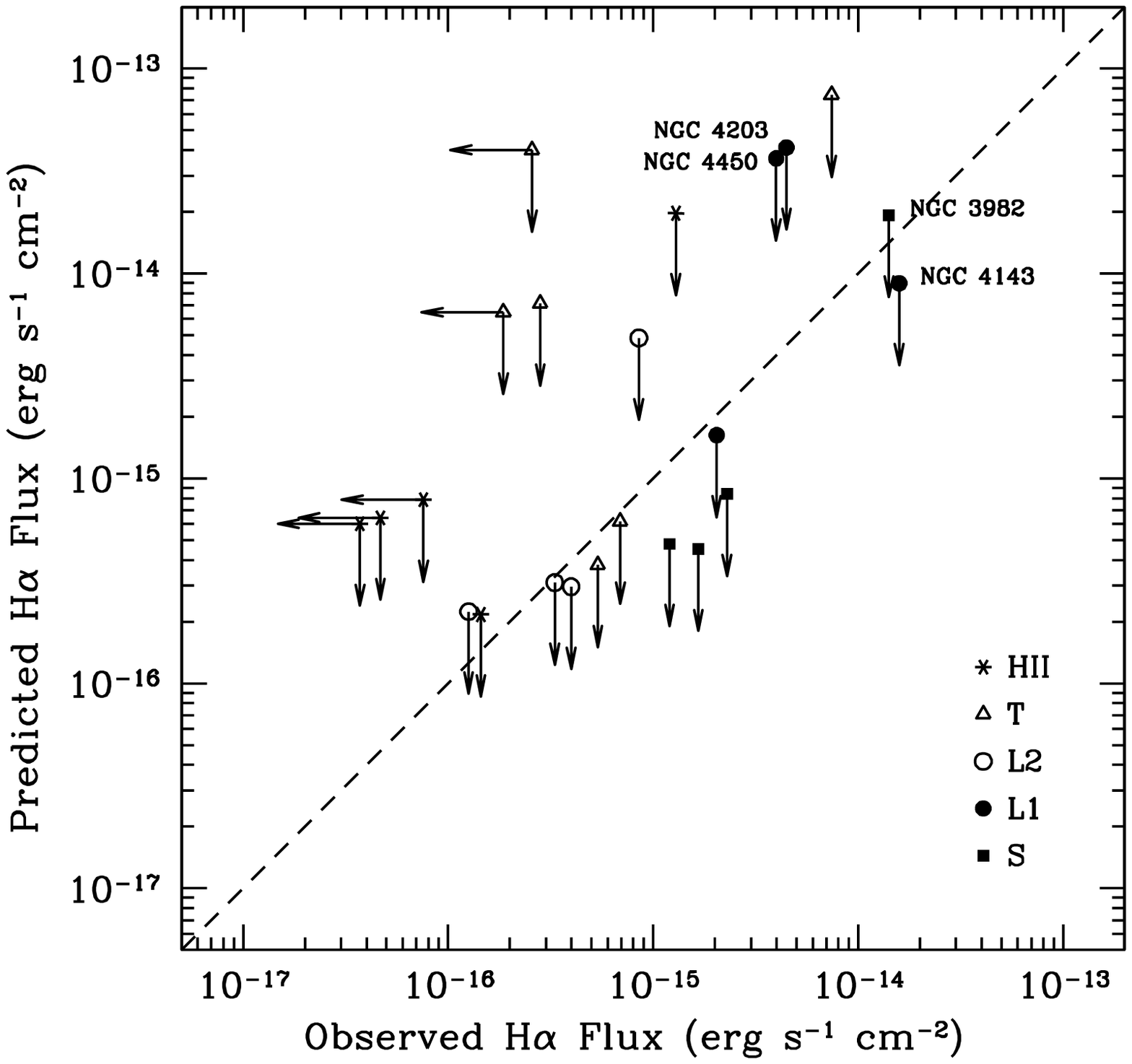}\caption{
 Comparison of the predicted H$\alpha$ fluxes for \ion{H}{2} regions
 powered by the maximum number of young stars indicated by our G430L
 spectra, and the observed H$\alpha$ emission in our G750M spectra.
 The symbol type denotes the ground-based, nuclear emission-line
 classification of our objects, while the {\it dashed line} shows the
 unity, consistency limit. Sources in which the stellar fluxes are
 probably overestimated due to contamination by an AGN continuum are
 labeled (NGC 3982, NGC~4143, NGC~4203, and NGC~4450). Correction of
 the observed H$\alpha$ emission for internal extinction inferred from
 the observed decrement of the Balmer emission lines will further
 shift most plotted points to the right.\label{fig:Pops_Ha_predvsobs}
 }\end{figure}

 \begin{figure}\plotone{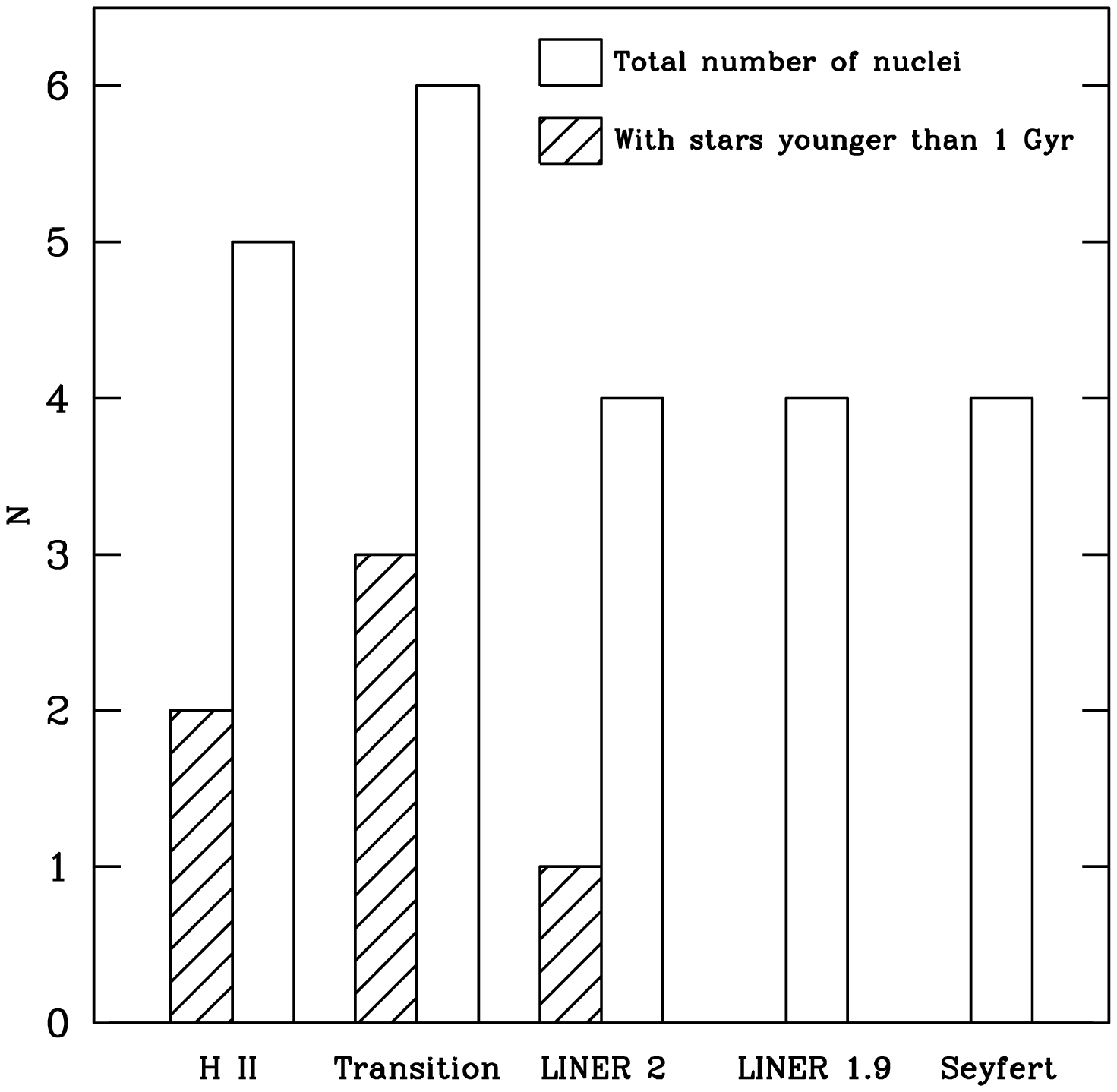}\caption{
 Incidence of stars younger that 1 Gyr in the surveyed galactic nuclei
 as a function of their ground-based spectral
 classification.\label{fig:Pops_with1GyrNuclClass}
 }\end{figure}

%%%%%%%%%%%%%%%%%%%%%%%%%%%%%%%%%%%%%%%%%%%%%%%
%
% TABLES
%
%%%%%%%%%%%%%%%%%%%%%%%%%%%%%%%%%%%%%%%%%%%%%%%

\clearpage
\begin{deluxetable}{lccccccccc}
\tabletypesize{\scriptsize}
%\footnotesize
\tablecaption{Basic Parameters of the Sample Galaxies \label{tab:Pops_GalSample}}
\tablewidth{0pt}
\tablehead{
\colhead{Galaxy} & \colhead{Hubble Type} & \colhead{$B_T$}            & \colhead{Spectral Class} & \colhead{$D$}    &
\colhead{$i$}    & \colhead{$A_{V,Gal}$} & \colhead{$\sigma_{\star}$} & \colhead{Ref.} & \colhead{Obs. Date} 
\\
\colhead{ }      & \colhead{ }           & \colhead{(mag)}            & \colhead{ }              & \colhead{(Mpc)}  & 
\colhead{(deg.)} & \colhead{(mag)}       & \colhead{(km s$^{-1}$)}    & \colhead{ } & \colhead{ }  
\\
\colhead{(1)} & \colhead{(2)} & \colhead{(3)} & \colhead{(4)} & 
\colhead{(5)} & \colhead{(6)} & \colhead{(7)} & \colhead{(8)} & \colhead{(9)} & \colhead{(10)}
}
\startdata
NGC~278  & SABb           & 11.47  & H     & 11.8 &  ---  & 0.46 &    ---     &  ---  & 10 Dec. 1998 \\
NGC~2787 & SB0$^+$        & 11.82  & L1.9  & 13.0 &    51 & 0.43 & $202 \pm  5$ & 1  & 05 Dec. 1998 \\
NGC~3351 & SBb            & 10.53  & H     &  8.1 &    48 & 0.09 & $101 \pm 16$ & 2  & 25 Dec. 1998 \\
NGC~3368 & SABab          & 10.11  & L2    &  8.1 &    47 & 0.08 & $135 \pm 10$ & 3  & 31 Oct. 1998 \\
NGC~3489 & SAB0$^+$       & 11.12  & T2/S2 &  6.4 &    56 & 0.05 & $112 \pm  3$ & 1  & 23 Jan. 1999 \\
NGC~3982 & SABb:          &  ---   & S1.9  & 17.0 &    30 & 0.05 & $ 73 \pm  4$ & 1  & 11 Apr. 1998 \\
NGC~3992 & SBbc           & 10.60  & T2:   & 17.0 &    53 & 0.09 & $140 \pm 20$ & 4  & 19 Feb. 1999 \\
NGC~4143 & SAB0$^{\circ}$ & 11.65  & L1.9  & 17.0 &    52 & 0.04 & $270 \pm 12$ & 5  & 20 Mar. 1999 \\
NGC~4203 & SAB0$^-$:      & 11.80  & L1.9  &  9.7 &    21 & 0.04 & $ 87 \pm  3$ & 6  & 18 Apr. 1999 \\
NGC~4245 & SB0/a          & 12.31  & H     &  9.7 &    41 & 0.07 &    ---     &  ---   & 18 Apr. 1999 \\
NGC~4314 & Sba            & 11.43  & L2    &  9.7 &    27 & 0.08 & $117 \pm  4$ & 1  & 20 Apr. 1999 \\
NGC~4321 & SABbc          & 10.05  & T2    & 16.8 &    32 & 0.08 & $ 92 \pm  4$ & 1  & 23 Apr. 1999 \\
NGC~4380 & SAb:?          & 12.66  & H     & 16.8 &    58 & 0.08 & $ 65 \pm 19$ & 3  & 25 Dec. 1999 \\
NGC~4435 & SB0$^{\circ}$  & 11.74  & T2/H: & 16.8 &    44 & 0.09 & $156 \pm  7$ & 7  & 29 Apr. 1999 \\
NGC~4450 & Sab            & 10.90  & L1.9  & 16.8 &    43 & 0.09 & $130 \pm 17$ & 2  & 31 Jan. 1999 \\
NGC~4459 & S0$^+$         & 11.32  & T2:   & 16.8 &    41 & 0.15 & $189 \pm 21$ & 1  & 23 Apr. 1999 \\
NGC~4477 & SB0:?          & 11.38  & S2    & 16.8 &    24 & 0.10 & $156 \pm 12$ & 8  & 23 Apr. 1999 \\
NGC~4501 & Sb             & 10.36  & S2    & 16.8 &    59 & 0.12 & $151 \pm 17$ & 9  & 26 Apr. 1999 \\
NGC~4548 & SBb            & 10.96  & L2    & 16.8 &    38 & 0.12 & $ 82 \pm  9$ & 10 & 26 Apr. 1999 \\
NGC~4596 & SB0$^+$        & 11.35  & L2::  & 16.8 &    43 & 0.07 & $154 \pm  5$ & 11 & 20 Dec. 1998 \\
NGC~4698 & Sab            & 11.46  & S2    & 16.8 &    53 & 0.08 & $134 \pm  6$ & 10 & 24 Nov. 1997 \\
NGC~4800 & Sb             & 12.30  & H     & 15.2 &    43 & 0.05 & $112 \pm  2$ & 1  & 03 Mar. 1999 \\
NGC~5055 & Sbc            &  9.31  & T2    &  7.2 &    56 & 0.06 & $118 \pm  4$ & 1  & 22 Mar. 1999 \\
\enddata
\tablecomments{
Col. (1): Galaxy name. Col. (2): Hubble type from de~Vaucouleurs et
al. (1991).  Col. (3): Total apparent $B$ magnitude from
de~Vaucouleurs et al. (1991).  Col. (4): Nuclear spectral class from
Ho et al. (1997), where H = \ion{H}{2} nucleus, L = LINER, S =
Seyfert, T = ``transition object'' (LINER/\ion{H}{2}), 1 = type~1, 2 =
type~2, and a fractional number between 1 and 2 denotes 
intermediate type; uncertain and highly uncertain classifications are
followed by a single and double colon, respectively.  Col. (5):
Distance, from Tully (1988), who assumes $H_0$ = 75 \kms\ Mpc$^{-1}$.
Col. (6): Galaxy inclination, from Ho et al. (1997).
Col. (7): Galactic foreground dust extinction, from Schlegel, Finkbeiner, \& Davis (1998).
Col. (8): Ground-based central stellar velocity dispersion
$\sigma_{\star}$.  Col. (9): Reference for $\sigma_{\star}$. 
Col. (10): UT observation date.\\
References. --- (1) Barth et al. 2002, red side; (2) Whitmore,
Schechter, \& Kirshner 1979; (3) H{\'e}raudeau et al. 1999; (4) Sarzi
\ea 2002; (5) Di~Nella \ea 1995; (6) Dalle Ore et al., 1991; (7)
Simien \& Prugniel 1997; (8) Jarvis \ea 1988; (9) H{\'e}raudeau \&
Simien 1998; (10) Corsini \ea 1999; (11) Kent 1990.
}
\end{deluxetable}

\clearpage
\begin{deluxetable}{lccc}
\tabletypesize{\footnotesize}
\tablecaption{Best Single-Starburst Models \label{tab:Pops_sage}}
\tablewidth{0pt}
\tablehead{
\colhead{Galaxy} & \colhead{$T_{best}$} & \colhead{$A_V$} & \colhead{$\chi^2_\nu$} \\
\colhead{ }      & \colhead{log(yr)}    & \colhead{(mag)}  & \colhead{ }              \\
\colhead{(1)}    & \colhead{(2)}        & \colhead{(3)}   & \colhead{(4)}
}
\startdata
NGC~278  & 8.50 & 0.4 &  1.85 \\
NGC~2787 & 10.0 & 0.0 &  2.53 \\
NGC~3351 & 9.25 & 0.0 &  1.34 \\
NGC~3368 & 9.00 & 0.5 &  1.88 \\
NGC~3489 & 9.00 & 0.3 &  3.14 \\
NGC~3982 & 9.25 & 0.0 &  2.49 \\
NGC~3992 & 10.0 & 0.8 &  1.52 \\
NGC~4143 & 9.50 & 0.1 &  3.50 \\
NGC~4203 & 9.25 & 0.0 & 17.58 \\
NGC~4245 & 10.0 & 0.0 &  1.43 \\
NGC~4314 & 10.0 & 0.3 &  1.29 \\
NGC~4321 & 9.00 & 0.0 &  7.99 \\
NGC~4380 & 10.0 & 0.5 &  0.72 \\
NGC~4435 & 9.75 & 0.4 &  1.32 \\
NGC~4450 & 9.25 & 0.1 &  8.46 \\
NGC~4459 & 10.0 & 0.5 &  3.63 \\
NGC~4477 & 10.0 & 0.6 &  1.76 \\
NGC~4501 & 10.0 & 1.0 &  1.80 \\
NGC~4548 & 10.0 & 0.6 &  1.20 \\
NGC~4596 & 10.0 & 0.2 &  1.50 \\
NGC~4698 & 10.0 & 0.2 &  1.60 \\
NGC~4800 & 10.0 & 0.1 &  1.56 \\
NGC~5055 & 9.50 & 0.0 & 11.08 \\
\enddata
\tablecomments{
Col. (1): Galaxy name. Col. (2)--(4): Age in Gyr of the best-fitting
Bruzual \& Charlot (2003) SSP model, required amount of internal extinction, and fit $\chi^2$
divided by the number of degrees of freedom, respectively.
}
\end{deluxetable}

\clearpage
\begin{deluxetable}{lccccc}
\tabletypesize{\footnotesize}
\tablecaption{Multiple-Starburst Models \label{tab:Pops_mage}}
\tablewidth{0pt}
\tablehead{
\colhead{Galaxy} & \colhead{$\overline{T}_{light}$} & \colhead{$A_V$} &
\colhead{$\chi^2_\nu$} & \colhead{L$_{\leq 1 {\rm Gyr}}$} & \colhead{M$_{\leq 1
{\rm Gyr}}$} \\
\colhead{ } & \colhead{log(yr)} & \colhead{(mag)} & \colhead{ } & \colhead{ {\footnotesize \%} } & \colhead{ {\footnotesize \%} }\\
\colhead{(1)} & \colhead{(2)} & \colhead{(3)} & \colhead{(4)} & \colhead{(5)} & \colhead{(6)}}
\startdata                      
NGC~278  & 8.50 & 0.4 & 1.86 & 100.00 & 100.00 \\ 
NGC~2787 & 10.0 & 0.0 & 2.56 &   ---  &   ---  \\ 
NGC~3351 & 9.39 & 0.4 & 0.79 &  45.16 &   1.31 \\ 
NGC~3368 & 9.51 & 0.5 & 1.24 &  22.66 &   1.37 \\ 
NGC~3489 & 9.49 & 0.0 & 1.98 &   8.95 &   0.94 \\ 
NGC~3982 & 9.29 & 0.3 & 1.46 &  38.70 &   0.42 \\ 
NGC~3992 & 10.0 & 0.8 & 1.53 &   ---  &   ---  \\ 
NGC~4143 & 9.91 & 0.2 & 2.34 &  19.02 &   0.15 \\ 
NGC~4203 & 9.68 & 0.5 & 4.39 &  45.24 &   0.32 \\ 
NGC~4245 & 9.99 & 0.0 & 1.42 &   1.64 &   0.06 \\ 
NGC~4314 & 10.0 & 0.3 & 1.29 &   ---  &   ---  \\ 
NGC~4321 & 9.01 & 0.5 & 1.58 &  51.44 &   2.44 \\ 
NGC~4380 & 10.0 & 0.4 & 0.72 &   ---  &   ---  \\ 
NGC~4435 & 9.95 & 0.4 & 1.23 &  11.98 &   0.18 \\ 
NGC~4450 & 9.79 & 0.4 & 2.33 &  37.82 &   0.26 \\ 
NGC~4459 & 10.0 & 0.5 & 3.61 &   ---  &   ---  \\ 
NGC~4477 & 10.0 & 0.6 & 1.77 &   ---  &   ---  \\ 
NGC~4501 & 10.0 & 1.0 & 1.81 &   ---  &   ---  \\ 
NGC~4548 & 10.0 & 0.6 & 1.20 &   ---  &   ---  \\ 
NGC~4596 & 10.0 & 0.2 & 1.50 &   ---  &   ---  \\ 
NGC~4698 & 10.0 & 0.3 & 1.60 &   ---  &   ---  \\ 
NGC~4800 & 9.99 & 0.1 & 1.45 &   2.10 &   0.07 \\ 
NGC~5055 & 9.87 & 0.1 &10.45 &  18.43 &   0.33 \\ 
\enddata 
\tablecomments{Col. (1): Galaxy name. Col. (2)--(4):
Luminosity-weighted mean age corresponding to the optimal combination
of templates listed in Table \ref{tab:Pops_mage_lweights}, required
amount of internal extinction, and fit $\chi^2$ divided by the number
of degrees of freedom, respectively. Col. (5)--(6): Cumulative light
and mass fraction from single-age B\&C models younger than 1 Gyr,
respectively. Only the stellar mass of the models is considered;
mass lost by stars during their evolution is excluded.}
\end{deluxetable}

\clearpage
\begin{deluxetable}{lccccccccccc}
\tablecolumns{12} 
\tabletypesize{\footnotesize}
\tablecaption{Multiple-Starburst Models \label{tab:Pops_mage_lweights}}
\tablewidth{0pt}
\tablehead{
\colhead{Galaxy} & 
\colhead{$l_{10^{6.0}}$}  &
\colhead{$l_{10^{6.5}}$}  &
\colhead{$l_{10^{7.0}}$}  &
\colhead{$l_{10^{7.5}}$}  &
\colhead{$l_{10^{8.0}}$}  &
\colhead{$l_{10^{8.5}}$}  &
\colhead{$l_{10^{9.0}}$}  &
\colhead{$l_{10^{9.25}}$} &
\colhead{$l_{10^{9.5}}$}  &
\colhead{$l_{10^{9.75}}$} &
\colhead{$l_{10^{10.0}}$} \\
\colhead{ } & 
\colhead{ } & \colhead{ } & \colhead{ } & \colhead{ } & \colhead{ } & \colhead{ } & \colhead{ } & \colhead{ } & \colhead{ } & \colhead{ } & \colhead{ } 
\\
\colhead{(1)} & 
\colhead{(2)} & \colhead{(3)} & \colhead{(4)} & \colhead{(5)} & \colhead{(6)} & \colhead{(7)} & \colhead{(8)} & \colhead{(9)} & \colhead{(10)} & \colhead{(11)} & \colhead{(12)} }
\startdata
NGC~278  &  ---  &  ---  &  ---  &  ---  &  ---  & 1.000 &  ---  &  ---  &  ---  &  ---  &  ---  \\
NGC~2787 &  ---  &  ---  &  ---  &  ---  &  ---  &  ---  &  ---  &  ---  &  ---  &  ---  & 1.000 \\
NGC~3351 &  ---  &  ---  & 0.253 &  ---  & 0.199 &  ---  &  ---  &  ---  & 0.450 &  ---  & 0.098 \\
NGC~3368 &  ---  &  ---  &  ---  &  ---  & 0.170 & 0.057 & 0.507 &  ---  &  ---  &  ---  & 0.266 \\
NGC~3489 &  ---  &  ---  &  ---  &  ---  & 0.006 & 0.084 & 0.614 &  ---  &  ---  & 0.121 & 0.175 \\
NGC~3982 &  ---  &  ---  & 0.387 &  ---  &  ---  &  ---  &  ---  &  ---  & 0.613 &  ---  &  ---  \\
NGC~3992 &  ---  &  ---  &  ---  &  ---  &  ---  &  ---  &  ---  &  ---  &  ---  &  ---  & 1.000 \\
NGC~4143 &  ---  & 0.010 &  ---  & 0.181 &  ---  &  ---  &  ---  &  ---  &  ---  &  ---  & 0.810 \\
NGC~4203 &  ---  & 0.252 &  ---  & 0.200 &  ---  &  ---  &  ---  &  ---  & 0.104 &  ---  & 0.444 \\
NGC~4245 &  ---  &  ---  &  ---  &  ---  &  ---  & 0.016 &  ---  &  ---  &  ---  &  ---  & 0.984 \\
NGC~4314 &  ---  &  ---  &  ---  &  ---  &  ---  &  ---  &  ---  &  ---  &  ---  &  ---  & 1.000 \\
NGC~4321 &  ---  &  ---  & 0.380 &  ---  & 0.134 &  ---  &  ---  & 0.376 & 0.110 &  ---  &  ---  \\
NGC~4380 &  ---  &  ---  &  ---  &  ---  &  ---  &  ---  &  ---  &  ---  &  ---  &  ---  & 1.000 \\
NGC~4435 & 0.007 &  ---  &  ---  &  ---  & 0.113 &  ---  &  ---  &  ---  &  ---  &  ---  & 0.880 \\
NGC~4450 &  ---  & 0.155 &  ---  & 0.223 &  ---  &  ---  &  ---  &  ---  &  ---  &  ---  & 0.622 \\
NGC~4459 &  ---  &  ---  &  ---  &  ---  &  ---  &  ---  &  ---  &  ---  &  ---  &  ---  & 1.000 \\
NGC~4477 &  ---  &  ---  &  ---  &  ---  &  ---  &  ---  &  ---  &  ---  &  ---  &  ---  & 1.000 \\
NGC~4501 &  ---  &  ---  &  ---  &  ---  &  ---  &  ---  &  ---  &  ---  &  ---  &  ---  & 1.000 \\
NGC~4548 &  ---  &  ---  &  ---  &  ---  &  ---  &  ---  &  ---  &  ---  &  ---  &  ---  & 1.000 \\
NGC~4596 &  ---  &  ---  &  ---  &  ---  &  ---  &  ---  &  ---  &  ---  &  ---  &  ---  & 1.000 \\
NGC~4698 &  ---  &  ---  &  ---  &  ---  &  ---  &  ---  &  ---  &  ---  &  ---  &  ---  & 1.000 \\
NGC~4800 &  ---  &  ---  &  ---  &  ---  &  ---  & 0.021 &  ---  &  ---  &  ---  &  ---  & 0.979 \\
NGC~5055 &  ---  &  ---  & 0.016 &  ---  & 0.168 &  ---  &  ---  &  ---  & 0.106 &  ---  & 0.710 \\

\enddata \tablecomments{ Col. (1): Galaxy name. Col. (2)--(11):
relative light weights of each Bruzual \& Charlot (2003) SSP
model in the best-fitting multiple starburst model for each nuclear
spectrum.}
\end{deluxetable}

\clearpage
\begin{deluxetable}{lccccc}
\tabletypesize{\footnotesize}
\tablecaption{Multiple-Starburst Models with $Z=2.5Z_{\odot}$\label{tab:Pops_mage_ss}}
\tablewidth{0pt}
\tablehead{
\colhead{Galaxy} & \colhead{$\overline{T}_{light}$} & \colhead{$A_V$} &
\colhead{$\chi^2_\nu$} & \colhead{L$_{\leq 1 {\rm Gyr}}$} & \colhead{M$_{\leq 1
{\rm Gyr}}$} \\
\colhead{ } & \colhead{log(yr)} & \colhead{(mag)} & \colhead{ } & \colhead{ {\footnotesize \%} } & \colhead{ {\footnotesize \%} }\\
\colhead{(1)} & \colhead{(2)} & \colhead{(3)} & \colhead{(4)} & \colhead{(5)} & \colhead{(6)}}
\startdata                                     
NGC~278  &   8.50 & 0.2 & 1.50 & 100.00 & 100.00 \\ 
NGC~2787 &   9.92 & 0.0 & 2.29 &  11.25 &   0.23 \\ 
NGC~3351 &   9.31 & 0.5 & 0.79 &  39.05 &   0.53 \\ 
NGC~3368 &   9.23 & 0.5 & 1.22 &  30.29 &   2.78 \\ 
NGC~3489 &   9.17 & 0.0 & 1.76 &  19.35 &   3.27 \\ 
NGC~3982 &   9.30 & 0.3 & 1.45 &  35.95 &   0.07 \\ 
NGC~3992 &   10.0 & 0.3 & 1.07 &  ---   &   ---  \\ 
NGC~4143 &   9.85 & 0.0 & 2.19 &  18.81 &   0.08 \\ 
NGC~4203 &   9.58 & 0.4 & 4.12 &  41.99 &   0.04 \\ 
NGC~4245 &   9.89 & 0.0 & 1.34 &  10.80 &   0.38 \\ 
NGC~4314 &   9.97 & 0.0 & 1.15 &   6.52 &   0.19 \\ 
NGC~4321 &   9.10 & 0.5 & 1.56 &  47.36 &   0.98 \\ 
NGC~4380 &   10.0 & 0.0 & 0.68 &  ---   &   ---  \\ 
NGC~4435 &   9.80 & 0.3 & 1.24 &  15.79 &   0.20 \\ 
NGC~4450 &   9.65 & 0.3 & 2.33 &  32.84 &   0.03 \\ 
NGC~4459 &   10.0 & 0.0 & 2.33 &   0.39 &   0.01 \\ 
NGC~4477 &   10.0 & 0.1 & 1.35 &  ---   &   ---  \\ 
NGC~4501 &   10.0 & 0.5 & 1.40 &  ---   &   ---  \\ 
NGC~4548 &   10.0 & 0.1 & 1.04 &  ---   &   ---  \\ 
NGC~4596 &   9.96 & 0.0 & 1.15 &   8.03 &   0.24 \\ 
NGC~4698 &   9.95 & 0.0 & 1.39 &   2.23 &   0.07 \\ 
NGC~4800 &   9.90 & 0.0 & 1.39 &   7.45 &   0.26 \\ 
NGC~5055 &   9.71 & 0.1 &10.30 &  13.62 &   0.14 \\ 
\enddata 
\tablecomments{Col. (1): Galaxy name. Col. (2)-(6):
Luminosity-weighted mean age corresponding to the optimal combination
of templates with super-solar metallicities, along with the
corresponding required amount of internal extinction, fit $\chi^2$
divided by the number of degrees of freedom, and cumulative light and
mass fraction from templates younger than 1 Gyr, respectively.}
\end{deluxetable}

\clearpage
\begin{deluxetable}{lcc}
\tabletypesize{\footnotesize}
\tablecaption{Maximum Predicted H$\alpha$ Powered by %\\ 
Young Stars \label{tab:Pops_ionizingfluxes}}
\tablewidth{0pt}
\tablehead{
\colhead{Galaxy} & \colhead{Predicted H$\alpha$ Flux}                        & \colhead{Observed H$\alpha$ Flux} \\
\colhead{ }      & \colhead{${\rm erg\; s^{-1} cm^{-2}}$} & \colhead{${\rm erg\; s^{-1} cm^{-2}}$} \\
\colhead{(1)}    & \colhead{(2)}                   & \colhead{(3)}}
\startdata
NGC~278  & $  7.9\times10^{-16} $ & $\leq 7.6\times10^{-17}\;\;\;\, $ \\
NGC~2787 & $  1.6\times10^{-15} $ & $     2.0\times10^{-15} $ \\
NGC~3351 & $  2.0\times10^{-14} $ & $     1.3\times10^{-15} $ \\
NGC~3368 & $  4.8\times10^{-15} $ & $     8.5\times10^{-16} $ \\
NGC~3489 & $  6.5\times10^{-15} $ & $\leq 1.9\times10^{-16}\;\;\;\, $ \\
NGC~3982 & $ (1.9\times10^{-14})$ & $     1.4\times10^{-14} $ \\
NGC~3992 & $  3.8\times10^{-16} $ & $     5.4\times10^{-16} $ \\
NGC~4143 & $ (9.0\times10^{-15})$ & $     1.5\times10^{-14} $ \\
NGC~4203 & $ (4.1\times10^{-14})$ & $     4.5\times10^{-15} $ \\
NGC~4245 & $  2.2\times10^{-16} $ & $     1.4\times10^{-16} $ \\
NGC~4314 & $  3.1\times10^{-16} $ & $     3.3\times10^{-16} $ \\
NGC~4321 & $  7.4\times10^{-14} $ & $     7.4\times10^{-15} $ \\
NGC~4380 & $  6.0\times10^{-16} $ & $\leq 3.7\times10^{-17}\;\;\;\, $ \\
NGC~4435 & $  7.1\times10^{-15} $ & $     2.8\times10^{-16} $ \\
NGC~4450 & $ (3.6\times10^{-14})$ & $     4.0\times10^{-15} $ \\
NGC~4459 & $  6.2\times10^{-16} $ & $     6.9\times10^{-16} $ \\
NGC~4477 & $  4.8\times10^{-16} $ & $     1.2\times10^{-15} $ \\
NGC~4501 & $  8.4\times10^{-16} $ & $     2.0\times10^{-15} $ \\
NGC~4548 & $  3.0\times10^{-16} $ & $     4.0\times10^{-16} $ \\
NGC~4596 & $  2.2\times10^{-16} $ & $     1.4\times10^{-16} $ \\
NGC~4698 & $  4.5\times10^{-16} $ & $     1.7\times10^{-15} $ \\
NGC~4800 & $  6.4\times10^{-16} $ & $\leq 4.7\times10^{-17}\;\;\;\, $ \\
NGC~5055 & $  4.0\times10^{-14} $ & $\leq 2.6\times10^{-16}\;\;\;\, $ \\
\enddata 

\tablecomments{Col. (1): Galaxy name. Col. (2): Predicted H$\alpha$
fluxes from the rate of ionizing photons that would be emitted by the
$3\sigma$ upper limit on 1-Myr-old stars obtained from our
spectra. Predictions that are questionable due to probable power-law
contamination of the continuum are indicated in parentheses. Col (3):
Observed H$\alpha$ emission in our G750M spectra (Shields \ea 2004). Upper
limits correspond to cases where no clear emission is detected.}

\end{deluxetable}

\end{document}